\documentclass[sigconf]{acmart}
\AtBeginDocument{%
  }

\copyrightyear{2026}
\acmYear{2026}
\setcopyright{cc}
\setcctype{by}
\acmConference[CHI '26]{Proceedings of the 2026 CHI Conference on Human Factors in Computing Systems}{April 13--17, 2026}{Barcelona, Spain}
\acmBooktitle{Proceedings of the 2026 CHI Conference on Human Factors in Computing Systems (CHI '26), April 13--17, 2026, Barcelona, Spain}
\acmPrice{}
\acmDOI{10.1145/3772318.3791114}
\acmISBN{979-8-4007-2278-3/2026/04}

\usepackage{pifont}
\newcommand{\cmark}{\ding{51}} % check mark
\newcommand{\xmark}{\ding{55}} % x mark

\usepackage{xcolor}
\usepackage{soul}
\newif\ifhighlightchanges

\newcommand{\newmarker}[1]{%
\ifhighlightchanges
\textcolor{black}{#1}%
\else
#1%
\fi
}

\newcommand{\deletemarker}[1]{%
\ifhighlightchanges
\else
\fi
}

\newcommand{\newnewmarker}[1]{
\ifhighlightchanges
\textcolor{black}{#1}
\else
#1%
\fi
}

\highlightchangestrue
\definecolor{mynewmarker}{rgb}{0,0,1}

\begin{document}

\title{The Eye–Head Mover Spectrum: Modelling Individual and Population Head Movement Tendencies in Virtual Reality}


\author{Jinghui Hu}
\orcid{0000-0002-3965-2474}
\email{j.hu23@lancaster.ac.uk}
\affiliation{%
    \institution{Lancaster University}
    \city{Lancaster}
    \country{United Kingdom}
}

\author{Ludwig Sidenmark}
\orcid{0000-0002-7965-0107}
\email{lsidenmark@dgp.toronto.edu}
\affiliation{%
    \institution{University of Toronto}
    \city{Toronto}
    \country{Canada}
}

\author{Hock Siang Lee}
\orcid{0000-0001-6263-6811}
\email{h.s.lee3@lancaster.ac.uk}
\affiliation{%
    \institution{Lancaster University}
    \city{Lancaster}
    \country{United Kingdom}
}

\author{Hans Gellersen}
\orcid{0000-0003-2233-2121}
\email{h.gellersen@lancaster.ac.uk}
\affiliation{%
  \institution{Lancaster University, United Kingdom}
  \city{Aarhus University}
  \country{Denmark}
}


% === Math helpers ===
\newcommand{\softplus}[1]{\log\!\bigl(1+e^{#1}\bigr)}
\newcommand{\ind}[1]{\mathbf{1}\!\left\{#1\right\}}

% === Model names ===
\newcommand{\ModelLinear}{\textsc{Linear }}
\newcommand{\ModelHinge}{\textsc{Hinge }}
\newcommand{\ModelSoftHinge}{\textsc{Soft Hinge }}

% === Symbols ===
\newcommand{\xamp}{x}                % gaze-shift amplitude (\textdegree{})
\newcommand{\yhead}{y}               % head contribution (deg or proportion)
\newcommand{\re}{r_{\mathrm{e}}}     % eye-only range
\newcommand{\gain}{g}                % head gain (slope)
\newcommand{\betaH}{\beta}           % asymptote (soft-hinge scale)
\newcommand{\tauH}{\tau}             % transition (threshold)
\newcommand{\sH}{s}                  % smoothness
\newcommand{\N}{N}

% === Fit metrics ===
\newcommand{\Rtwo}{\mathrm{R}^2}
\newcommand{\RMSE}{\mathrm{RMSE}}
\newcommand{\AIC}{\mathrm{AIC}}

\begin{abstract}
  People differ in how much they move their head versus their eyes when shifting gaze, yet such tendencies remain largely unexplored in HCI. We introduce head movement tendencies as a fundamental dimension of individual difference in VR and provide a quantitative account of their population-level distribution. Using a 360\textdegree{} video free-viewing dataset ($N=87$), we model head contributions to gaze shifts with a hinge-based parametric function, revealing a spectrum of strategies from eye-movers to head-movers. We then conduct a user study ($N=28$) combining 360\textdegree{} video viewing with a short controlled task using gaze targets. While parameter values differ across tasks, individuals show partial alignment in their relative positions within the population, indicating that tendencies are meaningful but shaped by context. Our findings establish head movement tendencies as an important concept for VR and highlight implications for adaptive systems such as foveated rendering, viewport alignment, and multi-user experience design.
\end{abstract}

% People differ in whether they prefer to move their head % or their eyes during gaze shifts. AHHH idk
% Individualistic tendencies for the head and eyes to contribute varying amounts during gaze shifts in VR have long been hypothesized in HCI, yet remained underexplored.



% \keywords{Do, Not, Us, This, Code, Put, the, Correct, Terms, for,
  % Your, Paper}


\maketitle

\section{Introduction}
Head and eye movements are tightly coupled during visual exploration. Because of common motor constraints, both eyes and head inevitably contribute to gaze shifts, particularly at larger amplitudes where head movement becomes necessary. Within these shared limits, however, individuals exhibit systematic behavioural heterogeneity in their eye–head coordination tendencies: some move their head even for smaller shifts, while others prefer to keep the head stable and rely more on their eyes. This difference was first introduced in psychology~\cite{fuller_head_1992}, referred to as head movers and non-head movers. Later studies demonstrated these tendencies were a continuum from head movers to non-head movers in both controlled laboratory experiments~\cite{thumser_idiosyncratic_2008} and real-world settings~\cite{thumser_eyehead_2009}.

Yet, the phenomenon remains underexplored in HCI and immersive technologies. Little work has systematically examined variation in head contribution to gaze shifts in VR and AR, where this behaviour is fundamental to the user experience. 
% It determines how users engage with virtual content in various ways.

% When exploring immersive scenes, greater head involvement leads to more viewpoint reorientation, while lower head involvement relies more on peripheral vision. 
This phenomenon determines how users engage with virtual content in various ways.
When exploring immersive scenes, greater head involvement leads to more viewpoint reorientation, while lower head keeps the head more static and instead relies on directing gaze to wider eccentricities. 
This difference directly shapes spatial awareness and comfort, and how people use gaze- or head-based interaction techniques, such as gaze-assisted pointing \cite{pfeuffer_gaze_2017} or gaze-plus-head selection \cite{hou2024gazeswitch}. 
Greater head involvement aligns the head quickly, and lower head involvement keeps actions more gaze-driven. In multi-user collaboration, greater head involvement produces overt orientation cues that are easy for others to interpret, while lower head involvement leads to subtler cues that make joint attention less transparent \cite{bovo_cone_2022,maloney_talking_2020}. This variation also affects the alignment of what collaborators actually see, as differences in head movement change how closely their viewpoints match when looking at the same content \cite{gottsacker_decoupled_2025,lee_patterns_2024}. Finally, at the system level, modelling variation in head contribution improves gaze prediction \cite{hu_sgaze_2019} and adaptive viewport streaming \cite{corbillon2017viewport, hosseini2016adaptive}, and can support more efficient foveated rendering \cite{patney2016towards,  krajancich2023towards,pan_head-eyek_2025}.

Eye and head movements are fundamental to VR/AR experience, shaping viewpoint orientation~\cite{lee2024snap}, visual access~\cite{zhao2019seeingvr,weir2023see}, perception~\cite{diemer2015impact} and various interaction techniques~\cite{hu2024skimr,hu2024lookup,hu2022evaluation}.
Recognising this variation opens opportunities to design VR/AR systems that are more adaptive, comfortable, and accessible. The essential first step toward such adaptation is to build a broader understanding of the phenomenon itself: \begin{itemize}
    \item RQ1: How does head contribution to gaze shifts vary between individuals?
    \item RQ2: How prevalent are different tendencies in the population?
    \item RQ3: Are individual tendencies in head contribution consistent across tasks?
\end{itemize}

Existing metrics are limited to address these questions. Prior approaches rely on relative ratios or arbitrary thresholds that are hard to interpret without a baseline~\cite{fuller_head_1992,thumser_idiosyncratic_2008,thumser_eyehead_2009}. At the individual level, they reduce behaviour to discrete and averaged values, missing how head contribution changes with amplitudes. At the population level, they categorise people into “head movers” and “non-head movers”, oversimplifying mixed strategies and assuming strict boundaries. Most studies have also been small in scale, leaving open whether these differences generalise to larger populations. HCI needs continuous, user-specific models that can be parameterised, compared across individuals, and integrated into adaptive systems.

In this paper, we take the fundamental step by introducing the concept of a continuous eye-head mover spectrum in HCI, which captures systematic behavioural heterogeneity in head contribution under shared biomechanical constraints.
% \newmarker{formalising earlier categorical observations into} 
% We model horizontal head contribution to gaze shifts as a continuous function of target eccentricity, providing fine-grained and interpretable profiles of individual strategies. 
We propose a user-specific model that characterises horizontal head contribution to gaze shifts as a continuous function of target eccentricity, providing fine-grained and interpretable profiles of individual strategies. This model is not only a method for analysis, but a representation of individual behaviours that can serve as a basis for relevant applications such as adaptive interfaces and system-level optimisations. 
Using a large open 360\textdegree{} free-viewing dataset (N=80) after quality check, we map the distribution of behaviours across a population, establishing the prevalence of head contribution variation in immersive viewing. Through a controlled user study (N=28), we further examine stability across tasks (abstract vs. free viewing), projecting participants back into the dataset distribution and analysing their joint placement across tasks.

\section{Related Work}
\subsection{Eye–Head Coordination in Gaze Shifts}
\newmarker{We first outline foundational physiological and behavioural principles of eye–head coordination, spanning oculomotor control, movement science, and perceptual psychology.}
Gaze shifts are produced by the combined contribution of eyes and the head. The eyes can rotate up to about 50\textdegree{} in any direction, but rotations beyond 30\textdegree{} relative to the head are rarely sustained due to reduced stability and comfort ~\cite{stahl_amplitude_1999,land_looking_2009}. The head extends this range, with typical horizontal rotations of 80–90\textdegree{} and vertical rotations of 60–70\textdegree{} ~\cite{ferrario_active_2002}. Our focus is on the horizontal dimension, where the eye and head together cover the eccentricities most relevant for immersive viewing in VR.

Amplitude is the main factor that determines how much the head contributes. Small horizontal shifts below 20\textdegree{} are usually made almost entirely with the eyes, while larger shifts increasingly involve the head~\cite{land_looking_2009, goossens_human_1997, freedman_coordination_2008}. Around 30\textdegree{}, the head already contributes about one third of the displacement, and with greater amplitudes, its share increases proportionally~\cite{saeb2011learning}. Although the eyes can theoretically rotate further, in practice head rotation becomes the natural way to reach wider angles while keeping the eyes near the centre. 

At the end of a shift, the eyes are typically closer to the central position relative to the head, whereas the head takes on a larger share of the eccentricity. This recentering is supported by the vestibulo-ocular reflex (VOR), which maintains fixation on the target during ongoing head motion~\cite{tweed_eye-head_1995,bartz1966eye}. The outcome is that with increasing amplitude, head contribution systematically grows while the eyes settle back toward the midline.

In this work, we restrict the analysis to horizontal gaze shifts within -50\textdegree{} to +50\textdegree{}. This range reflects the maximum functional rotation of the eyes~\cite{land_looking_2009} and captures the amplitudes where coordination with the head is most active. Within these limits, torso involvement is not normally required~\cite{sidenmark2019eye}, making it possible to isolate eye–head behaviour without additional confounding factors. The restriction also provides a consistent and interpretable basis for modelling amplitude-dependent contributions across participants.

\deletemarker{Our analysis builds on this fundamental relationship. By focusing on the start and end points of horizontal gaze shifts, we capture how head contribution scales with eccentricity and examine how these profiles vary between individuals. Beyond this shared amplitude effect, people differ in how readily they involve the head, a variation that has been described in terms of “head-movers” and “non-head movers.”}
\newmarker{Our analysis builds on this fundamental relationship. By focusing on the start and end points of gaze shifts, we capture how \textbf{horizontal} head contribution scales with eccentricity. We next review general computational models of eye–head coordination.}

\subsection{\newmarker{Models of Eye-Head Coordination}}

\newmarker{Early work in biomechanics treated coordination as an emergent property of the physical eye–head system. For example, models of smooth pursuit~\cite{iskander_simulating_2017} simulate orbital tissues, extraocular muscles and head–neck mechanics to show that head movement naturally increases once the eye approaches its comfortable rotation range. Rather than encoding coordination rules, the model numerically integrates the physical equations of motion for both the eye and head. Coordination therefore emerges from how these mechanical systems respond to shared pursuit commands.}

\newmarker{Complementary approaches derive coordination from optimal control principles. Saeb et al. propose a neural model that learns to generate coordinated eye–head saccades through incremental adaptation guided by a simple cost function. The model uses a feedforward spatiotemporal map to convert desired gaze shifts into motor commands for the eyes and head, and a local learning rule—derived via gradient descent—to minimise a cost composed of gaze error and control effort. After learning, the model reproduces key experimental phenomena of the eye-only saccades, realistic velocity profiles, increasing head contribution with gaze amplitude, VOR-like post-saccadic compensation, and dependence on initial eye position. }

\newmarker{
Later in virtual avatar animation, Pan et al.~\cite{pan_head-eyek_2025} focused on sequential gaze behaviour instead of isolated saccades. Given a sequence of gaze targets, their model computes the head trajectory by balancing dwell time effects, anticipation of future gaze and strain-reducing trade-offs. Head-EyeK formalises this as an optimisation problem that minimises the eye-strain term that pulls the head toward the current gaze target, and a locomotion effort term that favours minimal head movement unless the intended dwell is long. Dwell time is used to modulate the relative weighting of the energies, resulting in smaller head turns for brief fixations and larger ones for longer looks. 
% While Pan et al. model how coordination unfolds, they do not analyse how much individuals differ in the magnitude or onset of head involvement, nor how such tendencies are distributed across tasks or across the population.
% In related computer-vision work, Nonaka et al.~\cite{nonaka_dynamic_2022} model temporal eye–head–body dependencies to infer gaze direction from head and body trajectories.
}

\newmarker{Our model focuses on a different aspect of the problem: rather than simulating specific gaze shifts, we aim to capture the general individual tendency in head contribution while respecting the shared biomechanical constraints. In essence, we model the differences within the similarities: all individuals exhibit the same fundamental amplitude-dependent pattern, but they vary systematically in how readily and how strongly they recruit the head. Our parametric formulation captures this common shape while allowing flexible individual deviations, and the subsequent population analysis quantifies how these variations distribute across users. This is a variability described in terms of “head movers” and “non-head movers".}

\subsection{Head-Movers and Non-head Movers}
%\textcolor{blue}{TODO: explain the relative definitions of head movers to motivate the population distribution}
\begin{table*}[t]
\centering
\begin{tabular}{p{5.5cm}  p{11cm}}
\toprule
 \textbf{Name} & \textbf{Description} \\
\midrule

 Head Gain Ratio~\cite{fuller_head_1992} & Ratio of head movement amplitude to target amplitude (measured for 40\textdegree{} gaze shifts).\\
% Higher values (≈1.0) indicate head movers, lower values (≈0.3 or below) indicate non-movers. Beyond ~40\textdegree{} orbital eccentricity, head movements are mandatory, so this measure applies only within the discretionary range. 

 Eye-Only Range~\cite{oommen_influence_2004,oommen_overlapping_2005,thumser_idiosyncratic_2008} & Defines the range of initial target eccentricities relative to the head where gaze shifts are primarily accomplished by eye movements. Operationally, this is the range where the probability of an eye-only saccade exceeds 50\%. An eye-only saccade is defined as one in which any associated head movement does not exceed 10\% of the shift.  \\

 Slope of the Eye-Head Ranges~\cite{thumser_idiosyncratic_2008,oommen_overlapping_2005} & Defines how head amplitude grows linearly with target eccentricity. \\

 Combined Oculomotor Range~\cite{stahl_amplitude_1999,stahl_eye-head_2001,thumser_idiosyncratic_2008,thumser_eyehead_2009} & Quantifies the customary range of eye eccentricity or the angular span of the central 90\% of eye-in-head positions. \\

 Head-Only Range~\cite{oommen_influence_2004,oommen_overlapping_2005,thumser_idiosyncratic_2008} & Quantifies the customary range of head-on-torso eccentricity or the angular span of the central 90\% of head-on-torso positions. \\

 Gaze-Only Range~\cite{thumser_eyehead_2009} & Quantifies the preferred range of visual exploration by the angular span subtending the central 90\% of the distribution of gaze angles. \\

 Head Movement Proportion~\cite{thumser_eyehead_2009} & The percentage of the recording period during which the head is in motion, determined by head velocity thresholds (e.g., yaw head velocity exceeding 50\textdegree{}/s or 5\textdegree{}/s). \\

 Eye–Head Coordination Proportion~\cite{thumser_eyehead_2009} & The percentage of saccades that are associated with a head movement, often defined by a head velocity threshold (e.g., 50\textdegree{}/s) within a specific time window of saccade onset.\\

% Amplitude & Gaze Shift Amplitude & Mean size of total gaze shifts (eye + head) during saccades \\
\bottomrule
\end{tabular}
\caption{Some of the metrics used in prior psychology studies on eye–head coordination~\cite{fuller_head_1992, stahl_amplitude_1999, oommen_influence_2004,oommen_overlapping_2005,thumser_idiosyncratic_2008,thumser_eyehead_2009}.}
\label{tab:metrics}
\end{table*}

% explain the metrics used in the prior psychology paper and motivate our model, discuss the cons of the measures
% relativeness of the definitions and relativeness of the tasks
%\textcolor{green}{add neural gate paper, add owl paper}

Early work in psychology introduced the idea that people differ in how readily they move their heads when shifting gaze. \autoref{tab:metrics} summarises the main metrics introduced to capture these tendencies.

Fuller ~\cite{fuller_head_1992} was the first to define head movers and non-head movers. He introduced Head Gain, calculated as the ratio of head amplitude to target amplitude for 40° shifts. Participants with gains close to 1.0 were classified as head movers, while those below 0.3 were non-head movers. This definition was restricted to a single amplitude and categorical cut-offs, offering only a coarse distinction.

% Beyond 40\textdegree{} orbital eccentricity, head involvement was considered mandatory, so the measure was restricted to the discretionary range where either eyes or head could be used. 

Stahl~\cite{stahl_amplitude_1999} introduced the Eye-Only Range (EOR), defined as the span of target eccentricities where the probability of an eye-only saccade exceeds 50\%. In practice, an eye-only saccade is one where head movement contributes less than 10\% of the gaze shift. Beyond this, Eye–Head Ranges (EHR) describe the zone where head amplitude increases with eccentricity, summarised by the EHR slope. These measures provided a structured way to quantify how head involvement scales with amplitude. However, they still rely on fixed thresholds and reduce behaviour to single values. EOR width is sensitive to the threshold chosen, while EHR slope assumes a linear increase, missing nonlinear patterns and finer individual variation. Despite these limitations, they informed our work: we use EOR and EHR slope as baseline comparisons and extend the same principle into a continuous parametric model that avoids thresholds.

In a later study, Thumser et al.~\cite{thumser_idiosyncratic_2008} analysed head and eye use with new range-based measures such as the CHOR and GOR. They found that people (N=20) differed widely in these ranges, showing stable personal patterns rather than falling into neat “types.” These patterns were also consistent in the outdoor environment.

Thumser and Stahl ~\cite{thumser_eyehead_2009} extended controlled lab studies and showed that the eye–head coupling tendencies are consistent across real-world contexts such as sitting, riding, and walking with proportional metrics such as HMoP and EHCoP.

Together, this line of work established that head-movement propensity varies across individuals and contexts. Yet both the definitions and the metrics imposed categorical boundaries, relied on arbitrary thresholds, and reduced behaviour to single values. Moreover, prior studies were typically based on small samples (N < 20), limiting generalisability. Unlike prior small-N studies, we use a large open dataset (D-SAV360, N=87) to examine eye-head tendencies at scale (see \autoref{datasetSec}). The eye–head mover spectrum introduced here addresses these limitations by modelling head contribution as a smooth function of eccentricity at the individual level and as a continuous distribution at the population level.

% \textcolor{blue}{TODO: metrics used in Pshychology (table)}

% \subsection{Eye and Head Movement in HMDs}
% \subsection{Head and Gaze Dataset}

% \begin{itemize}
%     \item table of 360 video/eye head coordination datasets, availability online, gaze, head, number of participants
%     \item all the dataset would assume: Head is always upright (no roll/tilt); The head position is fixed at the centre of a unit sphere

% \end{itemize}

\subsection{Eye-Head Coordination in VR}
% VR compare to physical world 
% in VR, performing head movement increased the accuracy of spatial locolisation

Prior work has shown that coordination between gaze, head, and other effectors in XR is closely linked to effort, comfort, and task strategy, particularly during sustained or constrained interaction~\cite{10.1145/3706598.3713743, li2025quantifying, biener2024hold}. Head movements are generally more frequent in VR than in comparable physical settings\cite{pfeil_comparison_2018}, making eye–head coordination particularly important in immersive environments.

More recently, HCI research has begun to revisit these ideas. Sidenmark and Gellersen~\cite{sidenmark_eyehead_2019} found strong variability in VR: some participants frequently moved their heads even when unnecessary, while others relied almost exclusively on the eyes until ocular range was exceeded. 
% \deletemarker{Lee et al. use the terms “head movers” and “non-head movers” when analysing behaviour in gaze-based viewport control tasks.also reported clear individual differences in viewport alignment tasks, where head movers’ yaw scaled reliably with selection amplitude, while non-movers instead relied on snap-based techniques. These differences can also be interpreted as preferences for extrinsic, world-centred versus intrinsic, head/body-centred reference frames, which are especially relevant in VR where spatial frames structure perception and interaction.} 
\newmarker{Lee et al.~\cite{lee_patterns_2024} use the terms “head movers” and “non-head movers” when analysing behaviour in gaze-based viewport control tasks. However, these differences arise from \textit{strategic use of the interaction techniques} rather than from intrinsic eye–head coordination. Because all three techniques they tested (Controller Snap, Dwell Snap, Gaze Pursuit) allow viewport rotation without moving the head, some participants chose to rely on the technique in all conditions and showed almost no head movement, whereas others produced substantially higher cumulative head yaw even when the technique made head movement unnecessary. Thus, while this work demonstrates that variation exists, it does not analyse intrinsic coordination or model head contribution across gaze-shift amplitudes.}

% \newmarker{In contrast, our work examines coordination itself: we quantify how head contribution scales with gaze-shift amplitude, model the behaviour continuously for each participant, and analyse population structure across both free-viewing and controlled tasks. Our continuous spectrum therefore extends the strategy-based observations in Lee et al. by characterising the underlying eye–head movement tendencies directly.}

Other work has linked variability to perceptual and functional outcomes. Bayer et al.  \cite{bayer_active_2024} showed that participants who engaged the head during gaze shifts achieved more accurate spatial localisation. Pan et al. \cite{pan_head-eyek_2025} proposed Head-EyeK, a model learned in VR tasks, showing that head involvement depends not only on amplitude but also on dwell time and anticipation of upcoming targets. Applied studies further demonstrate that eye-head metrics carry information beyond gaze alone. Ries et al. \cite{ries_decoding_2025} found that including head features improved classification of task difficulty in VR search, while Zhao et al. \cite{zhao_coordauth_2025} showed that eye-head coordination can serve as an implicit biometric for hands-free authentication.

\newmarker{A parallel line of research develops computational models that use gaze and head dynamics to predict short-term behaviour. DGaze~\cite{hu_dgaze_2020} and FixationNet~\cite{hu_fixationnet_2021} forecast future gaze or fixation locations in dynamic VR scenes using combinations of saliency, motion, and head-movement features. Rolff et al. ~\cite{rolff_when_2022} predict saccade onset timing from gaze–head dynamics, and David-John et al.~\cite{david-john_towards_2021} infer users’ intent to interact from moment-to-moment gaze behaviour. While these models demonstrate the predictive value of gaze–head signals, they treat head movement primarily as an input feature rather than analysing underlying coordination behaviour.}

These studies confirm that head movement tendency in VR is variable, consequential, and practically relevant. Yet definitions remain categorical and no systematic account exists of how head contribution scales with amplitude or distributes across a population. Our work addresses this gap by introducing the eye–head mover spectrum, which models head contribution continuously at the individual level and reveals a continuous distribution of strategies at the population level.

% \subsection{\newmarker{Computational Models of Eye-Head Coordination}}

% \subsection{\newmarker{VR Eye-Head Datasets}}

\section{Open Dataset: 360\textdegree{} Free Viewing}
% need a large number of pars in the dataset to know prevalence...
\label{datasetSec}

To understand eye-head coordination at both individual and population levels, there are two requirements. First, we need a model that captures individual behaviour as a continuous function, rather than reducing it to discrete categories or summary ratios. Second, we need a method that allows comparison between individuals while preserving the full shape of these behaviours, so that differences are interpretable and their prevalence across a population can be assessed. Crucially, establishing prevalence further requires a large participant sample.

We therefore analyse the D-SAV360 dataset (N=87), which provides head and gaze recordings from a large and balanced group of participants freely viewing 360\textdegree{} videos in VR. We begin by fitting a continuous, user-specific model of head contribution to gaze shifts for each participant (see \autoref{ModelIndividuals}). We then apply functional PCA to compare these models across individuals to derive a population distribution that reveals both the range and prevalence of different strategies. (see \autoref{PopulationDistribution}).

% RQs:
% \begin{itemize}
%     \item Do individual differences of head movement tendencies exist in VR experiences like in real-world settings? (\ref{360})
%     \item How prevalent are these different strategies? (\ref{360})
%     % \item Are these strategies consistent for different eye movements? (\ref{nature})
%     % \item Are these strategies consistent with different task complexities? (\ref{nature})
%     % \item How can we model the continuous spectrum of individual differences in gaze coordination strategies across contexts?
%     \item 
% \end{itemize}
\subsection{Dataset Selection}

To investigate individual eye-head coordination tendencies at scale, we required a dataset that is both large and representative. Prior psychology studies have provided rich insights but typically involved fewer than 20 participants, limiting their ability to capture population-level variability. Recent efforts have produced open VR datasets, but they differ in size, content, and whether they include both head and eye tracking. \autoref{tab:datasets} summarises major candidates.

\begin{table*}[h]
\centering
\caption{Comparison of publicly available VR datasets with head and/or eye tracking.}
\label{tab:datasets}
\begin{tabular}{p{3cm}p{0.7cm}p{2.8cm}p{1cm}p{1cm}p{3.0cm}p{1cm}}
\toprule
\textbf{Dataset} & \textbf{N} & \textbf{Task/Content} & \textbf{Eye} & \textbf{Head} & \textbf{Length} & \textbf{Public} \\
\midrule
David et al.~\cite{david_dataset_2018} & 57 & 360° Videos & \cmark & \cmark & 19 videos, 20s each & \cmark \\
OpenNEEDS~\cite{emery_openneeds_2021} & 44 & Scene Exploration & \cmark & \cmark & 2 scenes, 5 min each & \xmark \\
EHTask~\cite{hu_ehtask_2023} & 30 & Free Viewing, Search & \cmark & \cmark & 4 tasks, 1–2 min & \cmark \\
Rubow et al. 2024~\cite{rubow_dataset_2024} & 25 & Tracking, Search & \cmark & \cmark & 4 tasks, 90s each & \cmark \\
PAV-SOD~\cite{zhang_pav-sod_2023} & 40 & 360° Video & \cmark & \xmark & 67 clips, $\sim$29.6s each & \cmark \\
D-SAV360~\cite{bernal-berdun_d-sav360_2023} & \textbf{87} & 360° Video & \cmark & \cmark & 85 videos, 30s each & \cmark \\
\bottomrule
\end{tabular}
\end{table*}

Among these, \textbf{D-SAV360} stands out as the most suitable foundation for our study. It is the largest publicly available dataset with both head and gaze tracking, covers diverse real-world 360° video content, and provides long continuous free-viewing data. Its scale and ecological validity closely match our research questions on natural eye-head coordination in VR. We therefore adopt D-SAV360 as the basis for our modelling, focusing on its monoscopic video condition and excluding stereoscopic blocks where not all participants contributed usable data.

\subsection{Dataset Description}
% \textcolor{blue}{TODO: Dataset selection; add a table of datasets}

We used the D-SAV360 dataset~\cite{bernal-berdun_d-sav360_2023}, which contains head and gaze recordings from 87 participants freely viewing dynamic 360\textdegree{} videos in VR. Data were captured with an HTC Vive Pro Eye, recording gaze at 120 Hz and head orientation at 90 Hz.

Participants viewed the videos in two blocks while standing: one block of 25 stereoscopic videos and one block of 30 monoscopic videos. Each video lasted 30 seconds and was presented with first-order ambisonic audio. 
\newmarker{To provide a clearer characterisation of the stimuli, we draw on the scene-level annotations available in D-SAV360. The 85 videos span a broad range of scene types, including indoor and outdoor environments, natural and urban settings, and both focused scenes with a single dominant event and more exploratory scenes containing multiple potential regions of interest. The dataset reports Spatial Information (SI) and Motion Vector (MV) values for each video, established metrics for quantifying spatial detail and motion magnitude in 360° content. As illustrated in Figure 2 of the dataset paper~\cite{bernal-berdun_d-sav360_2023}, the stimuli cover a wide SI–MV space, ranging from low-motion scenes to videos with substantial temporal change. In addition, the first-order ambisonic recordings are accompanied by audio energy maps (AEMs) that describe the spatial distribution and intensity of sound sources over time, supporting analysis of audio-driven saliency. Together, these descriptors provide a structured account of the visual and auditory variability present in the stimulus set.}

For our analysis, we focus on the monoscopic videos, as this block was viewed by more participants (including 16 who only completed monoscopic viewing) and provided a larger number of gaze shifts per participant. Together, this choice ensures broader coverage and more reliable modelling of head contribution.

% This dataset is well suited to our goals, as it offers fine-grained head and eye movement data across a large sample, enabling us to model individual head contribution to gaze shifts and derive a population-level distribution of strategies in immersive viewing.

% \begin{itemize}
%     \item preprocessing, selection of the participants/video types
%     \item metrics selections from prior paper
%     \item descriptive results compare to the reference points from prior brain study papers (use the same angle ranges instead of 360)
%     % \item (TODO)extract the same number of gaze shifts for each bin amplitude 
% \end{itemize}

\subsection{Data Preprocessing}
% Choices to fit the models: how to control for the free viewing task given the number of gaze shifts and gaze shifts distributions are not the same for each participant. 
\deletemarker{For preprocessing, we first selected valid trials from the monoscopic condition. Trials were retained if they contained sufficient samples, lasted close to the expected 30 seconds, and had matching head and gaze streams with enough shared timestamps. Head orientation was interpolated to gaze timestamps to ensure synchronised signals while preserving the temporal resolution of gaze data. Finally, participants with too few valid videos were excluded, leaving 80 participants for analysis.}

\subsubsection{\newmarker{Dataset Sanity Check}}
\newmarker{Before modelling individual behaviour, we performed a sanity check of the monoscopic portion of the D-SAV360 dataset to ensure that the publicly released recordings were complete, synchronised, and suitable for analysis. This involved two stages: checking the integrity of the eye–head data for each video, and then verifying the number of valid videos for each participant.}

\newmarker{In the video level checks, for each monoscopic video, we synchronised the gaze (120 Hz) and head-orientation (90 Hz) streams by interpolating head data to the gaze timestamps. A video was excluded if this alignment revealed discontinuities—such as long gaps, irregular sampling, or insufficient temporal overlap—indicating that the two streams could not be reliably matched. After alignment, we inspected the duration of the overlapping segment and retained videos only if this interval was close to the intended 30 seconds (> 25 s), ensuring that both signals captured the same viewing period.}

\newmarker{At the participant level, we check the number of valid videos per participant. According to the dataset specification, each participant should contribute 30 monoscopic videos. After applying the video-level criteria, we included only participants who retained the expected number of valid monoscopic recordings. This ensured that every included participant provided a complete and comparable dataset.}

\newmarker{This sanity check confirmed the internal consistency of the recordings and ensured that our analysis was based on synchronised and complete eye–head data. These exclusions arose solely from dataset quality issues and were independent of all subsequent modelling procedures. After this process, $N = 80$ participants remained.}

% criteria: first to eye and head data for each video
% interpolation of head data frequency and eye data frequecy -> fail shows the data continuity problem
% length of hte eye and head data in time, crop to match the timestamps
% check the matched data total time period for each video, if close to 30s (25s-25s), then keep this video data.

% Then to the participant level
% check the number of total valid monoscopic videos for each participant, if they end up having 30 valid videos as the paper suggested then keep the participant. 

% Therefore, this preprocessing step of removing partitcipants in the original dataset is merely the sanity check of the dataset quality is actually as stated in the paper and does not relate to any analysis method.

\subsubsection{Gaze Shift Detection}

We detected gaze shifts as all movements between fixations, rather than limiting analysis to saccades, in order to account for contributions of both the eyes and the head\newmarker{, which is defined in the same way as~\cite{kothari2020gaze}}. Horizontal gaze angles were first smoothed with a 1-Euro filter~\cite{casiez20121} (min-cutoff = 1.0, $\beta = 0.0$) to suppress noise while retaining rapid transitions. 
\newmarker{The thresholds were selected following the tuning procedure described in the original paper~\cite{casiez20121}: (i) set $\beta = 0$ and adjust the minimum cutoff while the signal is stationary or moving slowly; (ii) increase $\beta$ only if rapid movements exhibit noticeable latency. A 1 Hz minimum cutoff is also the recommended mid-range setting for stabilising noisy human motion signals while retaining responsiveness, and it effectively removes micro-jitter in the headset gaze stream. And $\beta$ was set conservatively as our aim was denoising without adaptation for speed. These parameter choices provide a stable gaze signal for fixation detection and gaze shifts extraction. } 
Instantaneous velocity was then derived from the filtered signals, and fixations were defined as periods where gaze speed fell below 15\textdegree{}/s for at least 60 ms. \newmarker{This aligns with findings from Agtzidis et al.~\cite{agtzidis2019360}, who showed that thresholds near 10\textdegree{}/s are effective for distinguishing fixations from pursuit in naturalistic 360° VR viewing; our slightly higher threshold provides a conservative criterion for our dataset.} Fixations were padded by 10 ms and merged if separated by gaps shorter than 20 ms to avoid spurious fragmentation. All intervals outside of fixations were labelled as gaze shifts, thereby capturing the full set of gaze reorientations arising from coordinated eye and head movements. As we focus on the eye and head coordination patterns, we also filtered out all the gaze shifts larger than 50\textdegree{}. We further filter out the outliers where head movements were even larger than the target eccentricities. \newmarker{Such cases typically reflect transient tracking artefacts or momentary overshoot rather than meaningful coordination behaviour. These outliers were very uncommon ($<2\%$ of detected gaze-shift samples per participant), and removing them did not materially change the fitted soft-hinge functions.}

We also focused on horizontal directions as in the prior work from Thumser and Stahl ~\cite{stahl_amplitude_1999,thumser_idiosyncratic_2008,thumser_eyehead_2009}. However, they focused on saccades that occurred while the head was stationary and then examined whether head movements followed. Their approach isolates saccades, whereas our definition considers all fixation-to-fixation transitions, including naturalistic reorientations where eyes and head act together under free-viewing conditions in VR.

\subsubsection{Symmetry}

We tested whether head movement responses were symmetric across leftward and rightward gaze shifts by mirroring left-side data and comparing binned movement amplitudes. Across participants, the median correlation between mirrored left and right responses was 0.980 (IQR: 0.963–0.988), and the median normalised difference was 6.9\% (IQR: 5.5–10.0\%). This indicates strong point symmetry in head movement patterns. Such symmetry is consistent with prior findings in head-eye coordination under symmetric target presentations ~\cite{freedman_coordination_2008, stahl_amplitude_1999}. \newmarker{This strong left–right correspondence confirms that mirroring does not alter the amplitude–contribution relationship and therefore cannot influence the later analysis of fitted curves or the fPCA results.}
% We therefore fit a symmetric model to reduce parameters, improve interpretability, and ensure robustness across participants.

We therefore symmetrised the data by reflecting eccentricities onto the positive axis. At low eccentricities, some head movements were occasionally directed opposite to the target. Since our goal is to capture head contribution to gaze shifts, we set these opposite-direction values to zero, ensuring that only head movements aligned with the target direction are treated as contributing.

% \subsection{Smooth Pursuit and Visual Search}
% \label{nature}
% % do both reflect the same latent strategy?
% %discover new subtypes: reveal previously unnoticed mixed strategies, inconsistencies, or transitions.
% % Psychological measures often use binary cutoffs (e.g., head moved >10%).
% % GPR metrics are continuous 

% use the new dataset to show that smooth pursuit and gaze shifts have similar/different head propensity model?

% use residuals from amplitude and predicted head movement to model eye movement

\subsection{Model Individuals}
\label{ModelIndividuals}
% Objectives of selecting and fitting a model for head movement and target amplitude:

% To develop a model that reliably and efficiently captures individual head movement strategies and describes the population distribution, including where an individual would sit in the population.

% even though the number of videos and the lengths of the videos are controlled, we can't control the number of gaze shifts each participant made.

% The distribution of the predictor (gaze shifts amplitude) can affect the model fitting.

% handling the uneven samplings for each participant
\begin{figure*}
    \centering   
    \includegraphics[width=0.9\linewidth]{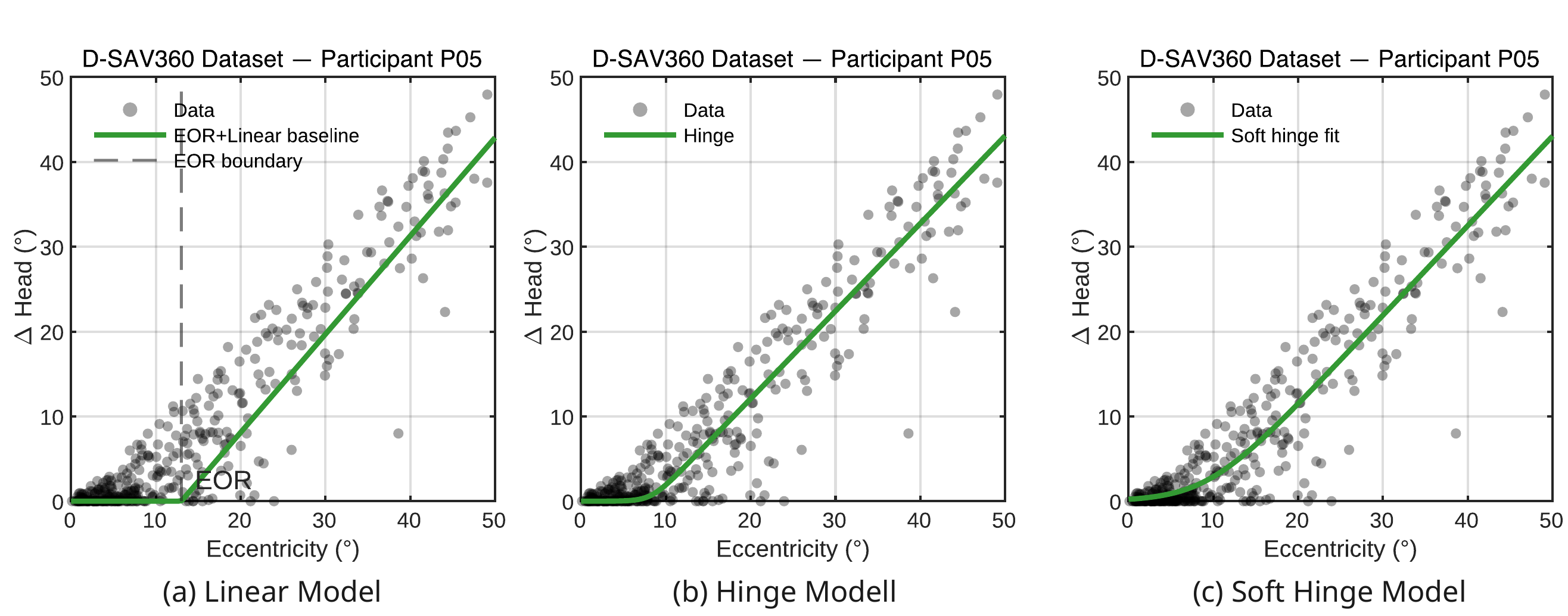}
    \caption{Example fits of three model formulations to one participant (P05).
Each plot shows horizontal head contribution ($\Delta$Head) against target eccentricity, with grey dots as data points. (a) Linear + EOR baseline assumes a fixed eye-only range followed by a linear increase. (b) Hinge model introduces a sharp breakpoint between eye- and head-dominant behaviour. (c) Soft hinge extends this with a smooth logistic transition, capturing both early and gradual head involvement.}
\Description{Three plots showing head contribution versus target eccentricity for one participant. The linear baseline has a sharp kink, the hinge has a sharp breakpoint estimated from data, and the soft hinge shows a smooth curve that follows the scatter most closely.}
    \label{fig:model_exp}
\end{figure*}

\begin{figure*}
    \centering   
    \includegraphics[width=0.7\linewidth]{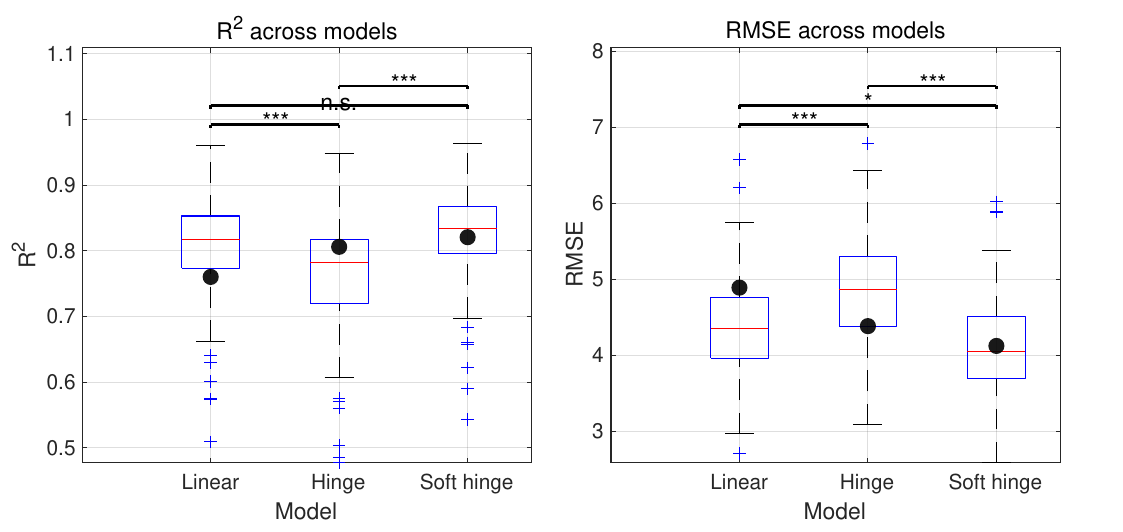}
    \caption{Comparison of model fits for gaze shift data.
Boxplots show the distribution of R² (left) and RMSE (right) across participants for three models: Linear baseline, two-parameter hinge, and three-parameter soft hinge. Boxes represent the interquartile range (IQR) with medians; black dots indicate group means. Significance bars show pairwise differences between models ($*p < .05$, $**p < .01$, $***p < .001$).}
\Description{Two boxplots comparing model performance. The soft hinge achieves the highest explained variance and lowest error across participants, outperforming the hinge and linear baselines.}

    \label{fig:model_fitness}
\end{figure*}

We aim to obtain a relatively robust and comparable description of head movement strategies for each participant. Specifically, we sought a model that (a) reliably captures how head contribution scales with gaze shift amplitude, even when data are sparse or unevenly distributed across eccentricities. The model, therefore, needs to represent this monotonic, amplitude-dependent structure; (b) provides a continuous measure of head movement tendencies, allowing fine-grained individual profiles; and (c) yields a uniform representation that supports meaningful comparisons between participants and across tasks.

\subsubsection{Model Selection}
\newnewmarker{}
We evaluated three parametric models, each inspired by prior psychological metrics (\autoref{tab:metrics}) and theoretical expectations of head contribution curves. These models represent different levels of flexibility while offering interpretable parameters. \autoref{fig:model_exp} shows the example fits of the three models. All models were defined over the same range of target eccentricities, 
\([0,50]^\circ\), corresponding to the theoretical maximum range of horizontal eye movements. 
Within this domain, we compared three formulations. 
\newmarker{All models take as input: }
\newmarker{$\xamp$: target eccentricity (\textdegree{}), defined as the horizontal angular displacement between two fixations.
$\yhead(\xamp)$: predicted horizontal head rotation amplitude (\textdegree{}) accompanying the gaze shift.}

\paragraph{Linear Model (a piecewise-linear fit with a breakpoint)}
A two-parameter formulation describing an eye-only range followed by a linear increase in head gain. This corresponds to early approximations of eye–head coordination using fixed ratios (see \autoref{tab:metrics}). 
Here, \(\text{EOR}\) and the corresponding EHRSlope (\(\gamma\)) 
are not free parameters but are pre-calculated for each participant 
following the definitions in the original work~\cite{stahl_amplitude_1999}.
\begin{equation}
\label{eq:linear}
\yhead(\xamp) \;=\gamma \cdot max(0,\xamp-\alpha)
\quad\text{(}\alpha=EOR,\ \gamma=EHR slope\text{)} , \quad x \in [0,50]^\circ
\end{equation}

\newmarker{Here, \(\xamp\) denotes target eccentricity (\textdegree{}), and \(\yhead(\xamp)\) denotes the predicted horizontal head-rotation amplitude (\textdegree{}). 
\(\alpha\) is the eye-only range (EOR), the eccentricity below which head movement is assumed to be zero, and 
\(\gamma\) is the linear rate at which head contribution increases beyond \(\alpha\). 
Both \(\alpha\) and \(\gamma\) are computed directly from each participant’s data and are not free model parameters.
}

\paragraph{Hinge Model} 
A step-like model that captures the transition from eye- to head-dominant behaviour with a threshold parameter.
% Two-parameter hinge (piecewise-linear with threshold)
\begin{equation}
\label{eq:hinge}
\yhead(\xamp) \;=\; \betaH \, \softplus{\xamp-\tauH}, \quad x \in [0,50]^\circ
\end{equation}

\newmarker{\(\xamp\) is the target eccentricity (\textdegree{}), and \(\yhead(\xamp)\) is the predicted head-rotation amplitude (\textdegree{}). 
\(\tauH\) is the threshold or ``knee'' eccentricity at which head involvement begins to increase noticeably.
\(\betaH\) is a scale factor controlling the overall magnitude of head movement across the curve.
This formulation produces a smooth hinge-like increase without assuming a fixed eye-only region.
}

\paragraph{Soft Hinge Model (a smooth logistic-shaped curve that approximates the hinge)}
An extension of the hinge with a smoothness term, allowing a gradual transition with eccentricity. This formulation provides interpretable parameters.
$\tauH$ marks the knee location (\textdegree), the eccentricity where head involvement starts to rise more steeply.
$\sH$ controls the softness of the hinge, i.e., how gradually the transition unfolds around $\tauH$.
$\betaH$ acts as a scale factor: together with 
$\sH$, it determines the asymptotic slope of the function for large eccentricities. Specifically, the slope tends to be \(\betaH/\sH\) as \(\xamp \to \infty\).
Intuitively, \(\tauH\) marks where the head starts to come in, \(\sH\) sets how gradually that transition happens, and \(\betaH\) scales how steeply head involvement increases once past the knee.

\begin{equation}
\label{eq:softhinge}
\yhead(\xamp) \;=\; \betaH \, \softplus{\frac{\xamp-\tauH}{\sH}}, \quad x \in [0,50]^\circ
\end{equation}

\newmarker{\(\xamp\) is target eccentricity (\textdegree{}), and \(\yhead(\xamp)\) is predicted head rotation (\textdegree{}).
\(\tauH\) is the knee location, 
\(\sH\) is the softness parameter, 
and \(\betaH\) determines the scale of head contribution.}
% motivations and benefits of using the hinge model
% s: the smoothness of head contribution

% a continuous measure of the head movement tendencies 

% a uniform measure to provide meaningful comparisons between individuals and tasks (comparing to grouping measures such as clustering.)

% also explains the 'mixed' group behaviours 

% a user profile which can also predict the head movements given a target eccentricity

% the distribution also allows for direct comparisons with no hard thresholds of labels. also shows the differences/similarities within the group members 

% a more fine-grained measure captures more information than the psych metrics

%ni hao, gambateh ᕙ(  •̀ ᗜ •́  )ᕗ - ruby

% Sparse, noisy, or uneven sampling → fitted curves help, because they smooth and standardise.

% \subsubsection{Hinge model}

% compare the baseline pshychology EOR and head gain model, the baseline psy model, soft hinge model of 2 parameters and 3 parameters.

% \subsection{Correlation of the three parameters }

\subsubsection{\newmarker{Fitting Procedure}}
\newmarker{For the hinge and soft-hinge models, we fitted the parameters directly to each participant's trial-level eye--head amplitudes using non-linear least-squares optimisation (MATLAB \texttt{lsqcurvefit}, trust-region-reflective), which minimised the sum of squared errors (SSE) between predicted and observed head contribution. The objective minimised the squared error between the model prediction and the observed head contribution for every gaze shift.} 

\newmarker{To ensure physiologically plausible solutions, we imposed minimal constraints on the model parameters. The slope parameter $\betaH$ was restricted to $[0,1]$, reflecting that head contribution cannot be negative or exceed the gaze amplitude. The transition parameter $\tauH$ and softness parameter $\sH$ were left free to capture individual differences in when and how smoothly head recruitment increases; in practice, the optimisation consistently converged to meaningful positive values. We employed a multi-start strategy (20 random initialisations) to mitigate sensitivity to local minima and retained the solution with the lowest SSE.}

\newmarker{The linear model was not fitted via optimisation. Following prior work, the breakpoint (EOR) and corresponding post-EOR slope were pre-computed for each participant and inserted directly into the model.}

\subsubsection{Model comparison (in-sample fitness).}
% \textcolor{blue}{TODO: model fit comparisons plots}

All three models were fit to each participant’s data ($N=80$), and their performance was evaluated in terms of explained variance ($\Rtwo$), error ($\RMSE$), and parsimony using Akaike Information Criterion ($\AIC$). Results are shown in \autoref{fig:model_fitness}.

$\Rtwo$ differed significantly between models (Kruskal-Wallis, $p<.001$): the \ModelSoftHinge  explained more variance than the \ModelHinge ($p<.001$) and was comparable to \ModelLinear ($p=.393$). $\RMSE$ (ANOVA, $p<.001$) favoured the \ModelSoftHinge over both alternatives ($p<.001$), with $\ModelLinear$ better than \ModelHinge.
$\AIC$ values were lowest for the \ModelSoftHinge ($M=959.38$, $SD=295.06$)) versus \ModelHinge ($M=998.69$, $SD=307.52$)) and \ModelLinear ($M=1077.97$, $SD=352.28$). Given its balance of fit, parsimony, and interpretability, we use the \ModelSoftHinge to define each participant’s profile ($\betaH$,$\tauH$,$\sH$) for subsequent population analyses.

\autoref{fig:model_exp} illustrates the behaviour of the three models 
on the same participant data. The \ModelLinear baseline produces a sharp 
kink at the fixed threshold, reflecting its discrete transition from eye-only 
to head-plus-eye movement. The \ModelHinge provides a similar piecewise form, 
but with the breakpoint estimated directly from the data, resulting in a 
better alignment with the observed head onset. The \ModelSoftHinge model further 
smooths this transition, capturing the gradual increase in head contribution 
without an artificial discontinuity. 
\newnewmarker{This smooth transition is also consistent with known eye–head coordination behaviour. Head recruitment does not occur at a strict threshold but increases progressively as ocular rotation approaches comfortable limits and coordination costs accumulate; small gaze shifts are primarily eye-driven, with head contribution increasing smoothly with amplitude.}
Visually, the soft-hinge curve follows 
the empirical scatter most closely across the whole eccentricity range, 
while the baselines underestimate head involvement.

In summary, while all three models captured the general increase of head contribution with eccentricity, the soft hinge consistently provided the most accurate and robust fits across participants. We therefore adopted the soft hinge model for subsequent analyses, using its parameters to define each participant’s individual head contribution profile. These profiles formed the basis for the population-level distribution described in the next section. In addition to analysis, the model also provides a compact representation of eye-head coordination that can be readily used for other applications.
% \textcolor{blue}{also the model can be used for application...}
\subsubsection{\newmarker{Sensitivity Check}}
\newmarker{To examine whether the soft-hinge fits depend on the specific fixation threshold, we repeated the preprocessing with thresholds of 10°/s and 20°/s and re-fitted the soft-hinge function for each participant. We then assessed robustness at the behavioural level by comparing the resulting head contribution curves across thresholds. Each fitted function was sampled on a uniform 0–50\textdegree{} eccentricity grid, and we computed the Pearson correlation between the curve obtained with the main threshold (15\textdegree{}/s) and those obtained with the alternative thresholds. This provides a direct measure of similarity in the overall shape of the reconstructed behavioural profile. Across participants, the correlations between different thresholds were consistently high (median r > 0.96). These results indicate that varying the fixation threshold within a plausible range affects only a narrow set of samples near the fixation-movement boundary and does not materially change the fitted soft-hinge functions individually.}

\subsection{Population Distribution}
\label{PopulationDistribution}
% We treat each participant’s soft hinge parameters as a compact behavioural fingerprint. These fingerprints form a distribution over a low-dimensional strategy space. This allows us to position new participants within the population and offers a way to personalise adaptive systems based on learned head movement strategies later.

% compare the rankings of psy clusters with the new method spectrum.
\begin{figure}[tb]
    \centering   \includegraphics[width=\linewidth]{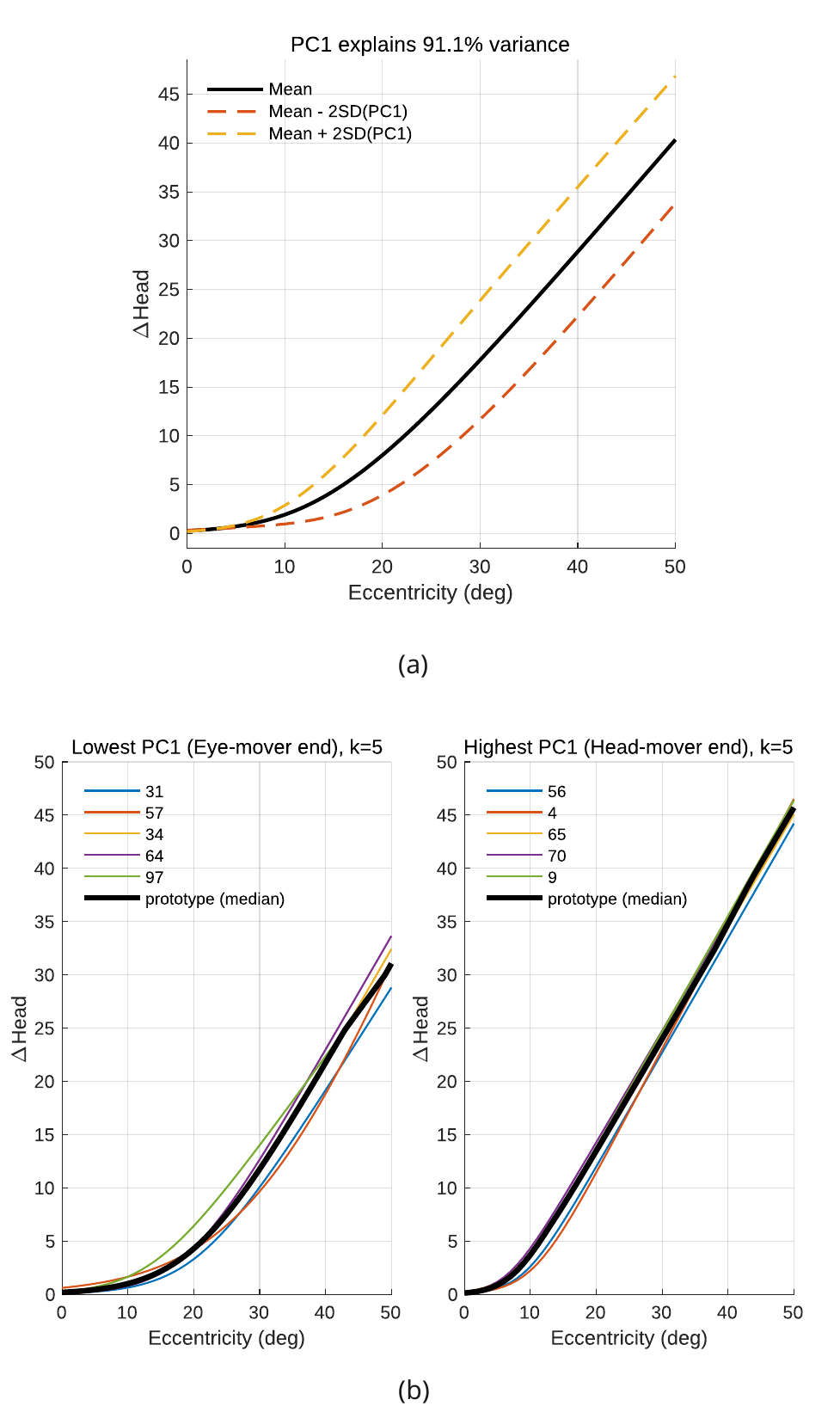}
    \caption{ Main spectrum of variation in head contribution revealed by fPCA across participants in the D-SAV360 dataset. fPCA identifies the dominant ways behavioural curves vary across participants.
(a) Reconstruction of the first fPCA component (91.1\% variance explained). The average curve is shown in black, with dashed curves indicating ±2 standard deviations along this component.
(b) Participants at the two ends of the same spectrum. Left: 5 participants with lower head contribution. Right: 5 participants with higher head contribution. Thin lines represent individual participants; the thick black line is the median curve within each group.}
\Description{Panel a shows the mean and ±2SD soft-hinge curves, with clear separation between low and high contributors. Panel b shows individual examples: five participants with low head contribution and five with high contribution, illustrating the spectrum.}

    \label{fig:fpca_360_pc1}
\end{figure}

\begin{figure}
    \centering   \includegraphics[width=0.7\linewidth]{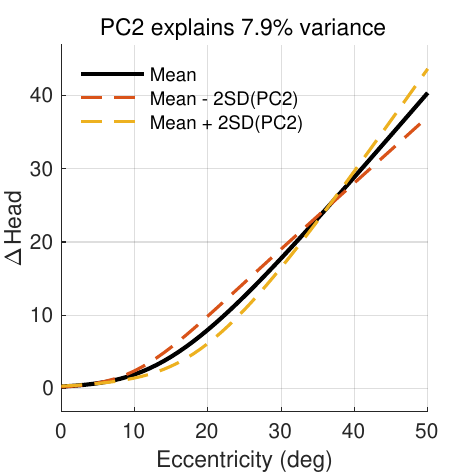}
    \caption{Secondary spectrum of variation identified by fPCA (D-SAV360 dataset).
The second fPCA component explains 7.8\% of the variance in the data. The average curve is shown in black, with dashed curves indicating ±2 standard deviations along this component. This secondary spectrum reflects subtler differences in head movement that go beyond the main continuum.}
\Description{A line plot showing the average curve and ±2SD variation for the second principal component. Variation is subtle, redistributing head contribution slightly between smaller and larger eccentricities.}

    \label{fig:fpca_360_pc2}
\end{figure}

After modelling individual head contribution profiles, we next examined how these behaviours are distributed across the population. Our goal was to assess how generalisable these tendencies are at scale, and to understand how much individuals differ in a large sample. To do this, we needed a representation that captures the variation across entire fitted curves, allows individuals to be compared within a common distribution, and still links each position in that distribution back to an interpretable curve.

\subsubsection{Method}

We applied functional principal component analysis (fPCA) directly to the set of fitted hinge curves. Intuitively, fPCA asks: across all participants, what is the dominant way their head-contribution curves differ and how does this behavioural heterogeneity organise along shared axes? Rather than focusing on individual parameters, it identifies shared axes of behavioural variation directly from the reconstructed coordination curves.
% \newmarker{Importantly, this analysis operates on the reconstructed behavioural curves rather than on the parameter vectors, ensuring that variability is captured in terms of the movement behaviour itself.}
Unlike standard PCA on parameters, fPCA considers each curve as a function and decomposes the collection into orthogonal components that describe the main ways in which the curves differ. Each curve was reconstructed from the fitted parameters [$\betaH$,$\tauH$,$\sH$] using the model (\autoref{eq:softhinge}). All curves were evaluated on a common eccentricity grid from 
0\textdegree{} to 50\textdegree{} in 1\textdegree{} steps. 
fPCA was then performed on the aligned curves. First, we centred the curves pointwise to obtain deviations from the sample mean function. We then computed the covariance across the grid and extracted orthogonal component functions that describe the dominant modes of variation. Each participant’s curve was projected onto these components to obtain scores that summarise their behaviour in a low-dimensional space. This representation makes it straightforward to position new curves relative to the existing distribution, while the component functions themselves indicate how variation along each axis changes the underlying head-contribution profile.

% To summarise population-level variation, we applied functional principal component analysis (fPCA) to the fitted curves. The first principal component (PC1) explained the majority of variance and primarily reflected whether participants tended to involve the head at lower or higher eccentricities. Projecting individuals onto this axis provided a convenient way to visualise the overall distribution of strategies in a single dimension.

\subsubsection{Results}

\begin{figure*}[htb]
    \centering   \includegraphics[width=0.8\linewidth]{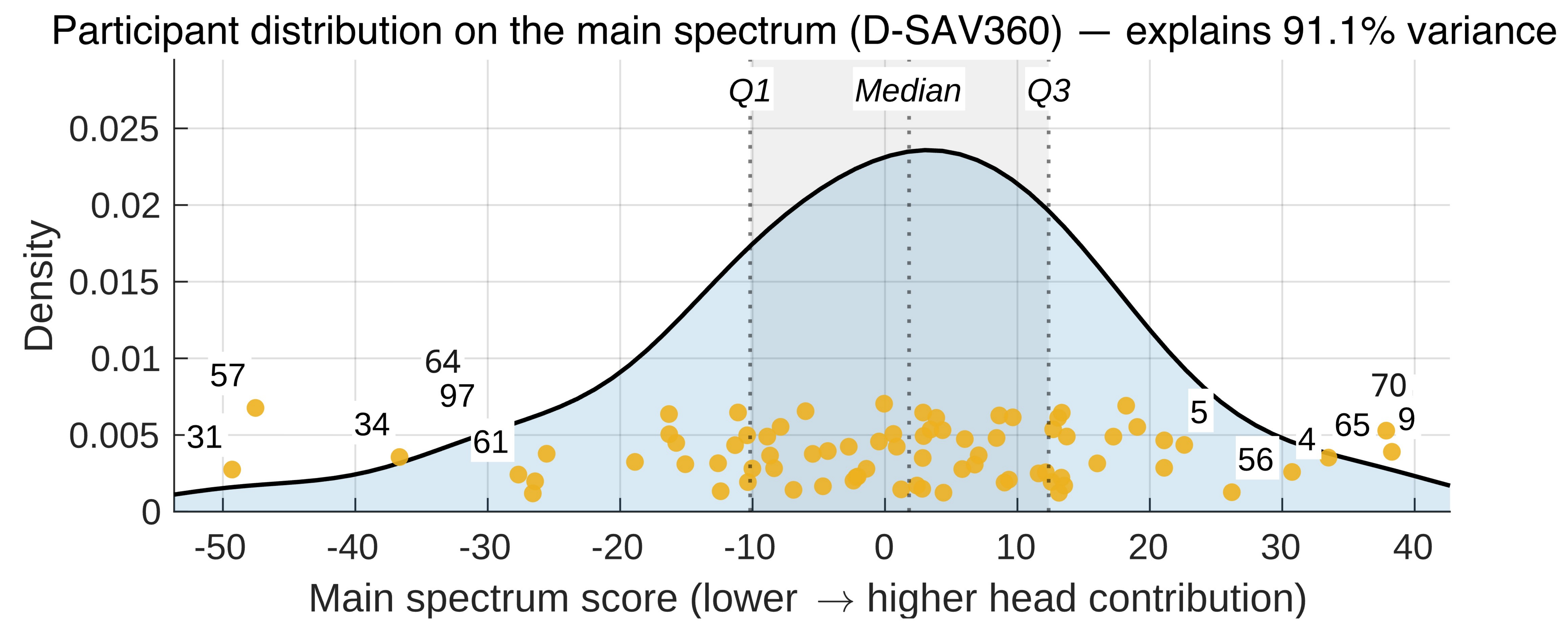}
    \caption{Participant distribution on the main spectrum of variation identified by fPCA (D-SAV360 dataset).
The x-axis shows each participant’s score on the first functional principal component (fPCA), which captures the continuum from lower to higher head contribution in gaze shifts. The smooth curve shows the overall distribution, dots mark individual participants, and the vertical lines indicate the 25th percentile (Q1), median, and 75th percentile (Q3). The numbers are the participants' IDs which mapped to the curves in \autoref{fig:fpca_360_pc1}(b).}
\Description{A density curve showing the distribution of participant scores along the main spectrum. Most cluster around the centre, with scattered individuals at both eye-leaning and head-leaning extremes.}

    \label{fig:fpca_360_distribution}
\end{figure*}

The functional PCA revealed that the first principal component (PC1) explained 91.1\% of the variance in head contribution, indicating that almost all individual differences could be captured along a single axis spanning from non-head-mover to head-mover tendencies (\autoref{fig:fpca_360_pc1}(a)).
This component reflects systematic differences in both the onset of head movement and the overall magnitude of contribution across eccentricities. The average curve (black line) rose smoothly with eccentricity, reaching about 38\textdegree{} $\Delta$Head at 50\textdegree{}. The ±2SD range (dashed curves) showed clear separation between lower and higher contributors: at 50\textdegree{}, head movements ranged from around 30\textdegree{} at the low end to 45\textdegree{} at the high end. Differences were already evident at smaller eccentricities. At 20\textdegree{}, head movements varied from only around 5\textdegree{} at the low end to nearly 15\textdegree{} at the high end—a threefold difference. At 30\textdegree{}, contributions diverged further, with lower contributors reaching around 12–15\textdegree{} compared to approximately 22–25\textdegree{} for higher contributors. From around 20\textdegree{} onward, there was a consistent difference of roughly 10\textdegree{} between the head-mover and eye-mover ends of the spectrum, a gap that was maintained up to 50\textdegree{}. Panel (b) illustrates these extremes with individual participants: those at the eye-mover end (e.g., P31, P57, P54, P64, P97) showed modest head involvement, with median curves reaching only about 31\textdegree{} at 50\textdegree{}, while those at the head-mover end (e.g., P56, P4, P65, P70, P9) exhibited strong head involvement, with medians approaching 45\textdegree{} at 50\textdegree{}.

The second component (PC2) explained only 7.8\% of the variance and reflected minor adjustments in curve shape, redistributing head contribution between smaller (5\textdegree{}–35\textdegree{}) and larger (35\textdegree{}–50\textdegree{}) eccentricities (\autoref{fig:fpca_360_pc2}). Compared to PC1, the effect of PC2 was subtle and did not substantially alter the overall profile of head involvement.

Participant scores on the main spectrum (see \autoref{fig:fpca_360_distribution}) covered a wide range from –49.30 (P31, eye-mover end) to +38.28 (P9, head-mover end) (Figure 3). The distribution was unimodal, with the median at +1.82 and an interquartile range from –10.18 (Q1) to +12.36 (Q3). Most participants clustered around the center, with the density curve peaking between 0 and +5, indicating that intermediate strategies were most common and that the overall balance was shifted slightly toward the head-mover side.

Despite this central tendency, the spread of scores was asymmetric. On the eye-mover side, participants were more widely scattered, with extreme values reaching –49.30 (P31) and –47.54 (P57), and additional individuals such as P34 (–36.66), P97 (–30.21), and P64 (–31.45) extending the lower tail. In contrast, the head-mover side was more compact, with fewer participants at the extreme and a maximum of +38.28 (P9), followed by P70 (+37.85), P65 (+33.49), P4 (+30.74), and P5 (+22.5).

In summary, almost all variation in head contribution was captured by a single component. Participants were broadly and continuously distributed along this axis, with intermediate strategies most common and a consistent separation of around 10\textdegree{} between extremes after 20\textdegree{}. These findings establish a continuous distribution of head-movement strategies across the population, rather than discrete subgroups.

% \subsection{Calibration}

% figure out the number of samples needed for each target eccentricity bin to saturate the model. 
% Then build a calibration application.

\section{User Study}
The large-scale dataset analysis revealed a continuous distribution of head movement tendencies in free-viewing of VR 360\textdegree{} videos. However, this dataset alone cannot answer whether these tendencies are stable across tasks or whether they are partly shaped by the context. To address this, we conducted a controlled user study combining free video viewing with an abstract target-eccentricity task. This allows us to test the stability and consistency of individual strategies.
\subsection{Task}
% motivate the study design: centrifugal targets to estimate the head movement propensity
% we assume this will be representative of an automatic head movement ....?
\subsubsection{Abstract Task}
\begin{figure}[tb]
    \centering
    \includegraphics[width=\linewidth]{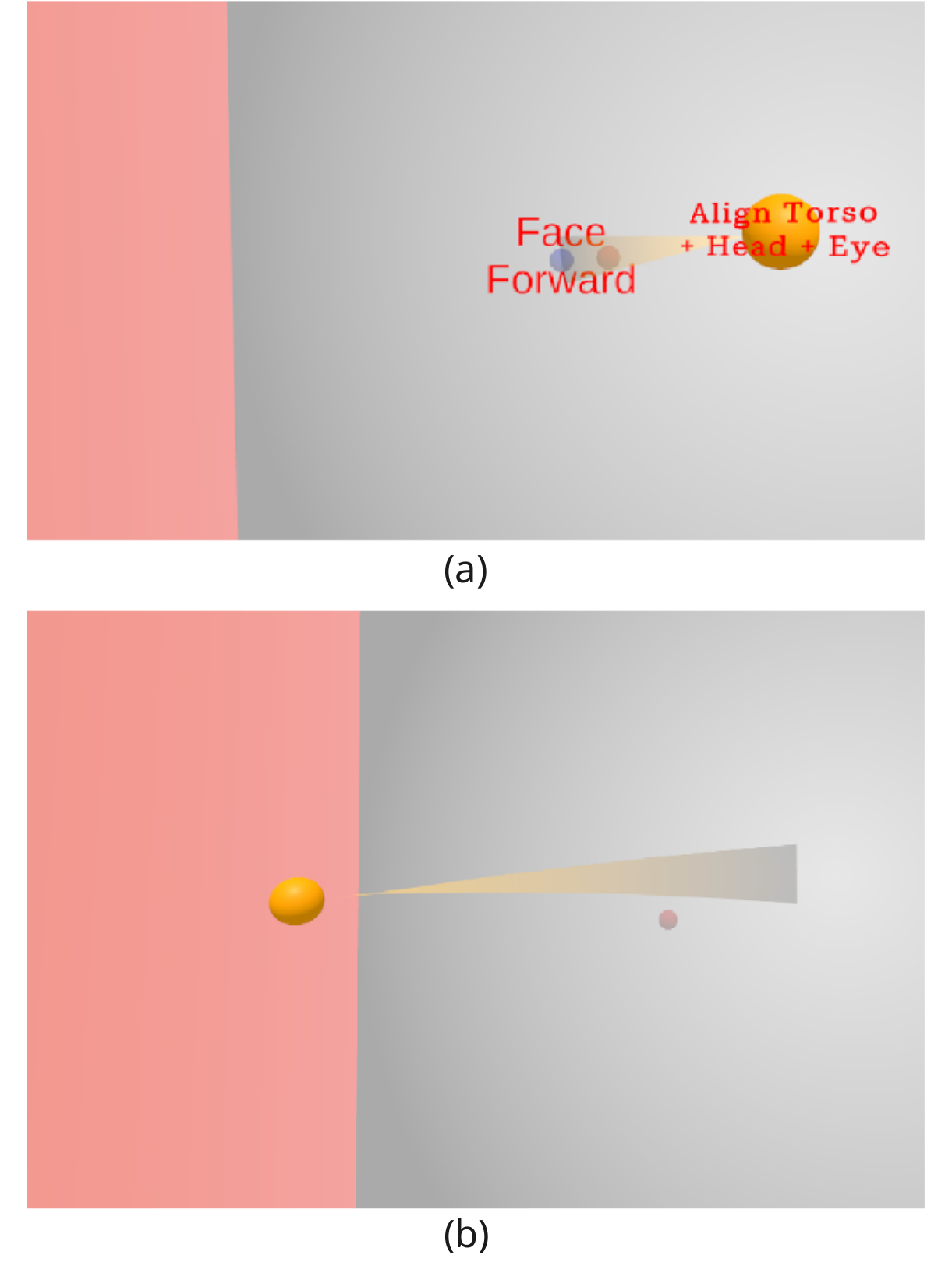}
    \caption{User study environment for the Abstract task. (a) At the start of each trial, participants aligned their eyes and head to a central target. (b) After alignment, a peripheral target appeared together with a visual guide indicating its direction. Participants shifted gaze toward the target and confirmed acquisition with a handheld controller.}
    \Description{Two panels showing the VR environment. Panel a: participant aligns eyes and head to a central target. Panel b: a peripheral target appears with a guiding arrow, prompting a gaze shift.}

    \label{fig:environ}
\end{figure}

The user study included a controlled abstract task designed to elicit gaze shifts under standardised starting conditions. The experimental design was inspired by prior work on eye–head coordination that presented targets arranged on a hemisphere and required participants to initiate shifts from a central fixation ~\cite{sidenmark_eyehead_2019}. We adapted this paradigm to VR by placing gaze targets along the horizontal meridian around the participant (\autoref{fig:environ}).

At the beginning of each trial, participants aligned their torso, head, and eyes to a central sphere target. Colour feedback indicated correct alignment. Once aligned, participants pressed a controller button to initiate the trial, upon which the central target disappeared and a new peripheral target appeared.

Targets were positioned in bins spanning from –50\textdegree{} to +50\textdegree{} in 5\textdegree{} increments. Within each bin, the exact target position was generated randomly, ensuring a balanced but non-repetitive sampling of eccentricities. Each participant completed 20 repetitions per bin, and the sequence of targets was randomised to prevent order effects and anticipation. In total, this resulted in 400 trials per participant.

Participants were instructed to shift their gaze towards the target as quickly and accurately as possible, using either their eyes, head, or both at their discretion. A visual guide indicated the direction of the target. Each trial ended when the participant’s gaze cursor was within the target sphere and confirmed by pressing the controller button, at which point the central alignment target reappeared.

\subsubsection{Video Task}
We replicated the video free-viewing procedure of the original D-SAV360 dataset in a shortened form. Each participant viewed a block of 20 monoscopic 360\textdegree{} videos, selected from the dataset based on those eliciting the highest number of gaze shifts across participants. This ensured the block was representative while aligning with our dataset analysis, which also focused on the monoscopic condition. The reduced block length kept the session duration comparable to the abstract task.

As in the original dataset, each video started from a controlled initial posture. Instead of locating a red cube in a black room (dataset procedure), participants in our study used the central sphere alignment: they aligned head and eyes to a central target and confirmed with a controller button. This adaptation preserved the controlled start while keeping the procedure consistent with the abstract task. Once aligned, the video played automatically for 30 seconds with ambisonic audio.

As in the original dataset protocol, participants were asked to watch the videos naturally, without any explicit instructions about how to move their head or eyes. To maintain engagement and detect non-compliant behaviour, we included the same four-alternative forced-choice sentinel question \newmarker{from the dataset paper~\cite{bernal-berdun_d-sav360_2023}} at the end of the video block, requiring participants to answer a simple content-related question:
\newmarker{\textit{At what time of the day is the video recorded? (i) The video is recorded during the day (ii) The video is recorded at night (iii) It is not possible to know, the video is recorded indoors (iv) I don’t know}}

\subsection{Procedure}
Upon arrival, participants received written information about the study, provided informed consent, and completed a demographic form along with a short pre-experiment Simulator Sickness Questionnaire (SSQ). The experimenter then explained the two tasks, after which participants were fitted with the Meta Quest Pro HMD.

All tasks were performed in a standing position, consistent with the original dataset protocol. The study followed a fixed order of Abstract task first and then Video task, with each task consisting of two blocks separated by optional short breaks. In addition, participants were asked to take off the HMD and rest in-between tasks until ready to continue if needed.

Before starting each task and again before each subset of ten videos, the built-in eye-tracking calibration was performed. Every abstract trial and video began with the same central-sphere alignment, in which participants aligned their head and eyes to a central target and confirmed with the controller to start. Abstract task blocks lasted about five minutes. Each video block contained 10 monoscopic 30-second clips (5 minutes in total, excluding alignment).

At the end of each block, participants answered a simple four-alternative forced-choice sentinel question. After completing both tasks, they filled out the post-study SSQ (identical to the pre-study version) and a short subjective questionnaire. The latter asked about: (i) awareness of their head movements during the Abstract task (7-point Likert, 1 = Not at all, 7 = Extremely), (ii) awareness during the Video task (same scale), (iii) preferred way of shifting gaze (eyes, head, or both), and (iv) whether their strategy differed between tasks (yes / no / don’t know).

The sentinel questions and SSQ responses were used to detect non-compliant participation. Data from participants who failed these checks were excluded prior to analysis. The entire study session lasted approximately 35 minutes.

% \textcolor{blue}{TODO: add study environment pic}

\subsection{Apparatus}

The study was implemented in Unity 2021.3.27f1 and run on a computer with an Intel Core i7-12700 CPU, 16 GB RAM, and an NVIDIA GeForce RTX 3070 Ti GPU. A Meta Quest Pro VR HMD was connected via Quest Link. The headset provided a 111.24\textdegree{} diagonal field of view, 1800 × 1920 pixel resolution per eye, and a 90 Hz refresh rate.

\subsection{Participants}
We recruited 28 participants (aged 22–54 years, M = 31.0, SD = 8.5). All 28 participants passed the sentinel questions and showed no abnormal increase in simulator sickness scores. 11 participants identified as female and 17 as male. All had normal or corrected-to-normal vision: one participant wore contact lenses, 18 (64\%) used glasses, and 9 (32\%) had uncorrected normal vision. In terms of VR experience, 16 participants (57\%) had used VR five times or fewer, including one participant who had never tried VR. Five participants (17\%) reported using VR often, and six (21\%) reported very frequent use. Regarding eye-tracking experience, 20 participants (70\%) had used eye-tracking devices fewer than five times, including three who had never used them before.

\subsection{Results}
All 28 participants' data were included as they passed the sentinel questions and showed no abnormal increase in sickness scores. Data were preprocessed with the same pipeline as the 360\textdegree{} dataset: gaze and head signals were synchronised, gaze shifts were extracted from fixation-to-fixation transitions and horizontal components were isolated. 
For the Abstract task, we analysed gaze and head orientations recorded at the time participants confirmed target acquisition with the controller. For both tasks, we removed the outlier gaze shifts where the head contributions were larger than the target eccentricities.

\subsubsection{Trial Effect (Abstract Task)}
\begin{figure}[tb]
    \centering
    \includegraphics[width=\linewidth]{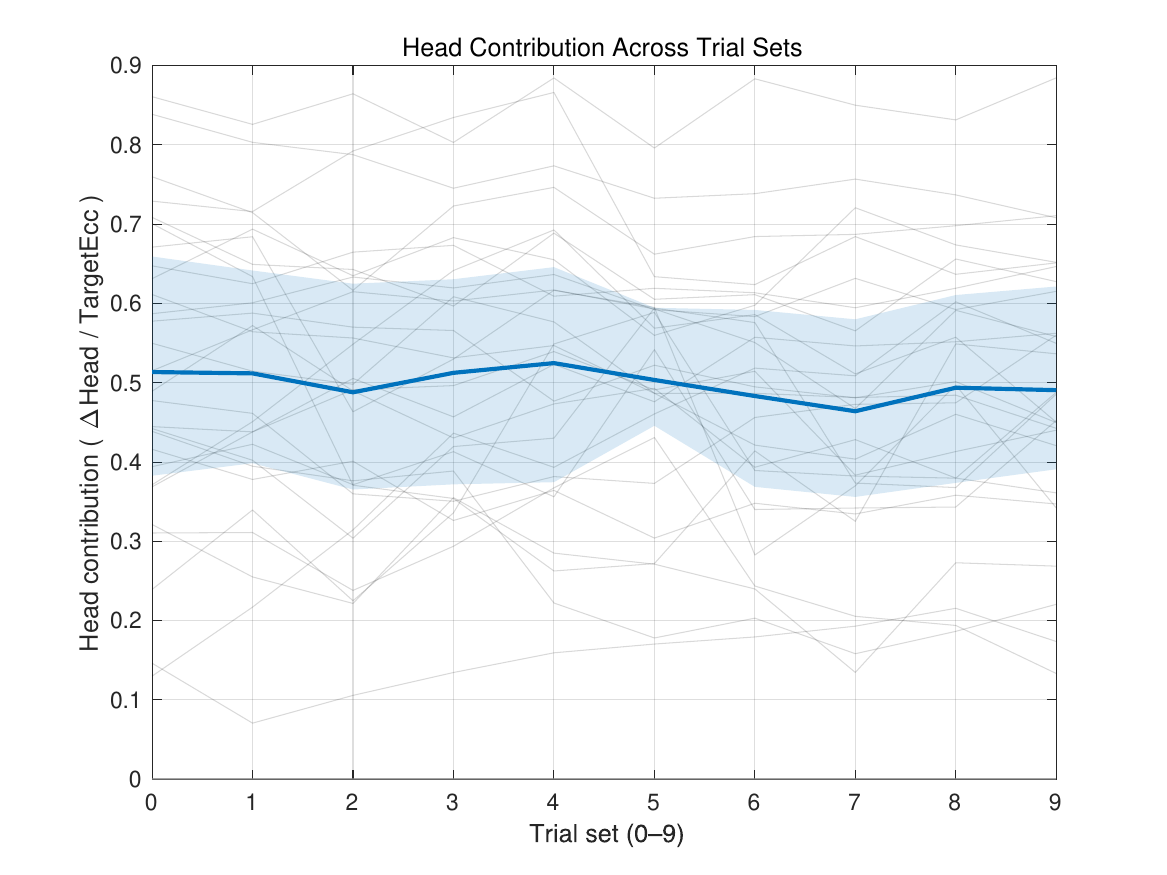}
    \caption{Learning effect of head contribution across trial sets. Each line represents the head contribution to gaze shifts, defined as $\Delta$Head / TargetEccentricity, averaged per trial set. Faint thin lines show individual participants, while the bold line indicates the group average across participants. The shaded band marks the interquartile range (25th–75th percentile).}
    \label{fig:HeadContributionTrialSet}
    \Description{Line plot showing head contribution over 10 trial sets. Individual thin lines represent participants, and the thick bold line shows group average. There is no trend over time, indicating stable strategies.}
\end{figure}

We examined whether head contribution ($\Delta$Head / target eccentricity) changed systematically across the ten trial sets. A repeated-measures ANOVA revealed no significant main effect of trial set ($F(9,243) = 1.51$, $p = .144$).

To compare early and late performance directly, we contrasted the first and final trial sets. Head contribution was 0.514 ($SD = 0.193$) in the first set and 0.491 ($SD = 0.173$) in the last set ($N = 28$). Normality of the paired differences was satisfied ($p = .233$), and a paired t-test confirmed no significant change ($t(27) = 0.81$, $p = .423$).

Hence, participants’ eye-head coordination strategy in the Abstract task remained stable across repetitions and all trial sets data can be used together for individual model fitting. 

\subsubsection{Model Fit}

\begin{figure}[tb]
    \centering
    \includegraphics[width=\linewidth]{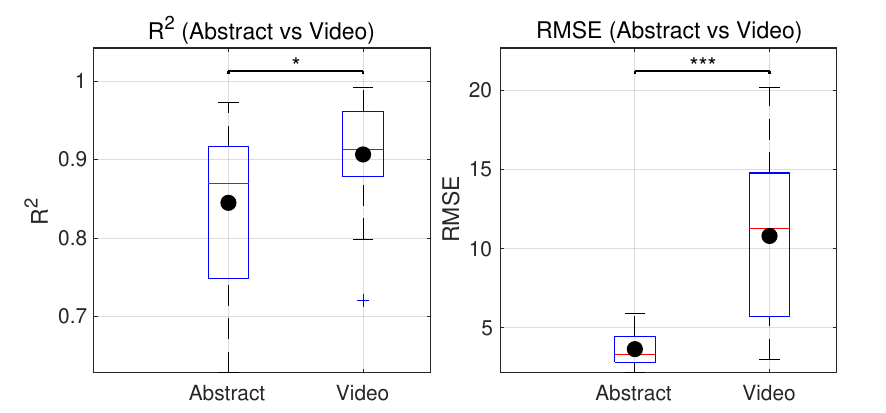}
    \caption{Comparison of model fit between Abstract and Video tasks.
Boxplots show $\Rtwo$ (left) and $\RMSE$ (right) across participants. Boxes indicate the interquartile range (IQR) with medians, and black dots mark group means. Significance bars report results of paired tests ($***p < .001$,  $*p < .05$), testing whether fitness differed between Abstract and Video tasks.}
\Description{Two boxplots comparing R² and RMSE across Abstract and Video tasks. Both have high R², but the Video task shows higher error due to larger, more variable gaze shifts.}
    \label{fig:TaskModelFitness}
\end{figure}

\begin{figure*}[htb]
    \centering   \includegraphics[width=0.8\linewidth]{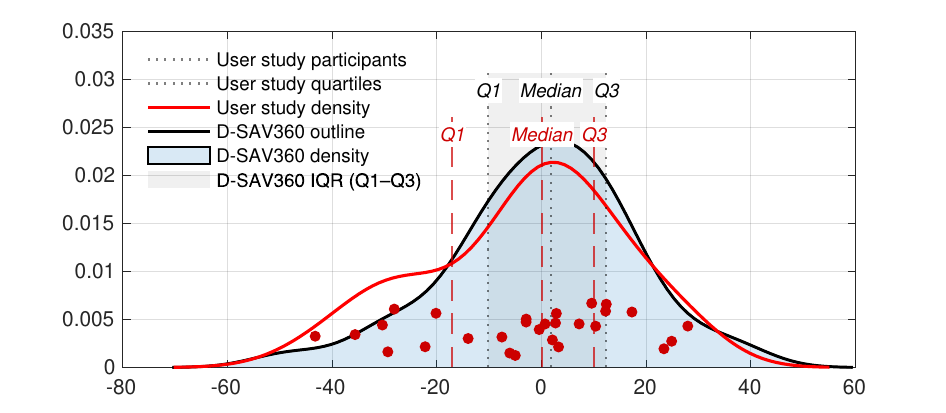}
    \caption{User study participants (video task) projected onto the D-SAV360 spectrum.
The black curve shows the distribution of PC1 scores in the D-SAV360 dataset, with shaded area indicating the interquartile range (IQR) and dotted lines marking dataset quartiles (Q1, Median, Q3). The red curve shows the distribution of user study participants, with dashed lines marking their quartiles. Red dots indicate individual participants. }
\Description{Overlaid density plots showing user study participants’ spectrum scores compared to the larger dataset. Both distributions align closely, with user study participants spanning the full range.}

    \label{fig:fpca_us_distribution}
\end{figure*}

We applied the same preprocessing procedure to the Video task data as in the dataset analysis, including the identical method for extracting gaze shifts. For the Abstract task, we instead used the gaze and head directions recorded at the moment when the controller was pressed while the gaze rested on the target; these data points formed the basis for analysis. In both tasks, we fit the soft-hinge model (\autoref{eq:softhinge}), selected from the dataset analysis, separately to the Abstract and Video data.

Model fits showed consistently high explained variance across tasks (\autoref{fig:TaskModelFitness}). In the Abstract task, mean $\Rtwo$ was 0.845 ($SD = 0.10$, $N = 28$), while the Video task achieved an even higher $\Rtwo$ of 0.907 ($SD = 0.065$). A paired $t$-test confirmed this difference as significant ($t(27) = 2.63$, $p = .014$). High $\Rtwo$ values in both conditions indicate that the soft-hinge model captured the overall structure of eye-head coordination reliably.

Residual errors, however, differed strongly between tasks. The Abstract task produced a mean $\RMSE$ of 3.66\textdegree{} ($SD = 1.08$), compared to 10.81\textdegree{} ($SD = 4.88$) in the Video task, a highly significant difference ($t(27) = 7.50$, $p < .001$). The lower $\RMSE$ in the Abstract task reflects precise fits with small residual errors, whereas the higher $\RMSE$ in the Video task reflects larger absolute deviations due to the greater magnitude and variability of gaze shifts during free viewing.

For reference, in the larger 360\textdegree{} video dataset ($N = 80$), $\RMSE$ values were typically lower, closer to those observed in the Abstract task. The higher $\RMSE$ in the lab Video task likely stems from two factors: fewer trials per participant, producing noisier individual fits, and the inherently larger, more variable gaze shifts elicited by naturalistic video viewing. Overall, the Video task fits remained statistically strong (high $\Rtwo$) but naturally exhibited higher absolute error magnitudes compared to both the Abstract task and the large-scale dataset.

Overall, the soft-hinge model generalises well across both controlled and naturalistic tasks, while highlighting task-dependent scaling of residual error.

\subsubsection{Projection of Video Task Fits onto Dataset Spectrum}

To compare the user study with the larger 360\textdegree{} dataset, we projected participants’ soft-hinge parameters from the Video task onto the principal component axis (PC1) derived from the dataset (\autoref{fig:fpca_us_distribution}). This places participants on the same behavioural spectrum that spans from eye-mover to head-mover strategies.

Participant scores covered a broad range within the D-SAV360 spectrum, spanning from –43.231 to +27.970, with quartiles at –17.075 (Q1), 0.133 (median), and +9.984 (Q3). The distribution was slightly skewed toward lower head contribution (skewness = –0.376), and showed a single peak around the central range.

All participants fell within the full spectrum defined by the D-SAV360 dataset (–49.302 to +38.278). At the extremes, participant P1 was located near the eye-mover end (PC1 = –43.231, 3.0th percentile of the dataset), while participant P20 was located near the head-mover end (PC1 = +27.970, 95.5th percentile).

Compared to the dataset baseline, the user study distribution (red curve in \autoref{fig:fpca_us_distribution}) overlapped closely with the D-SAV360 distribution (black curve in \autoref{fig:fpca_us_distribution}). Both medians were near zero, indicating similar central tendencies. The main difference was a slightly greater spread toward the eye-mover side (negative PC1 values) in the user study, suggesting a mild overrepresentation of lower head-contribution strategies relative to the larger dataset.

\subsubsection{Joint fPCA for Abstract vs. Video}

\begin{figure*}[tbh]
    \centering   \includegraphics[width=0.6\linewidth]{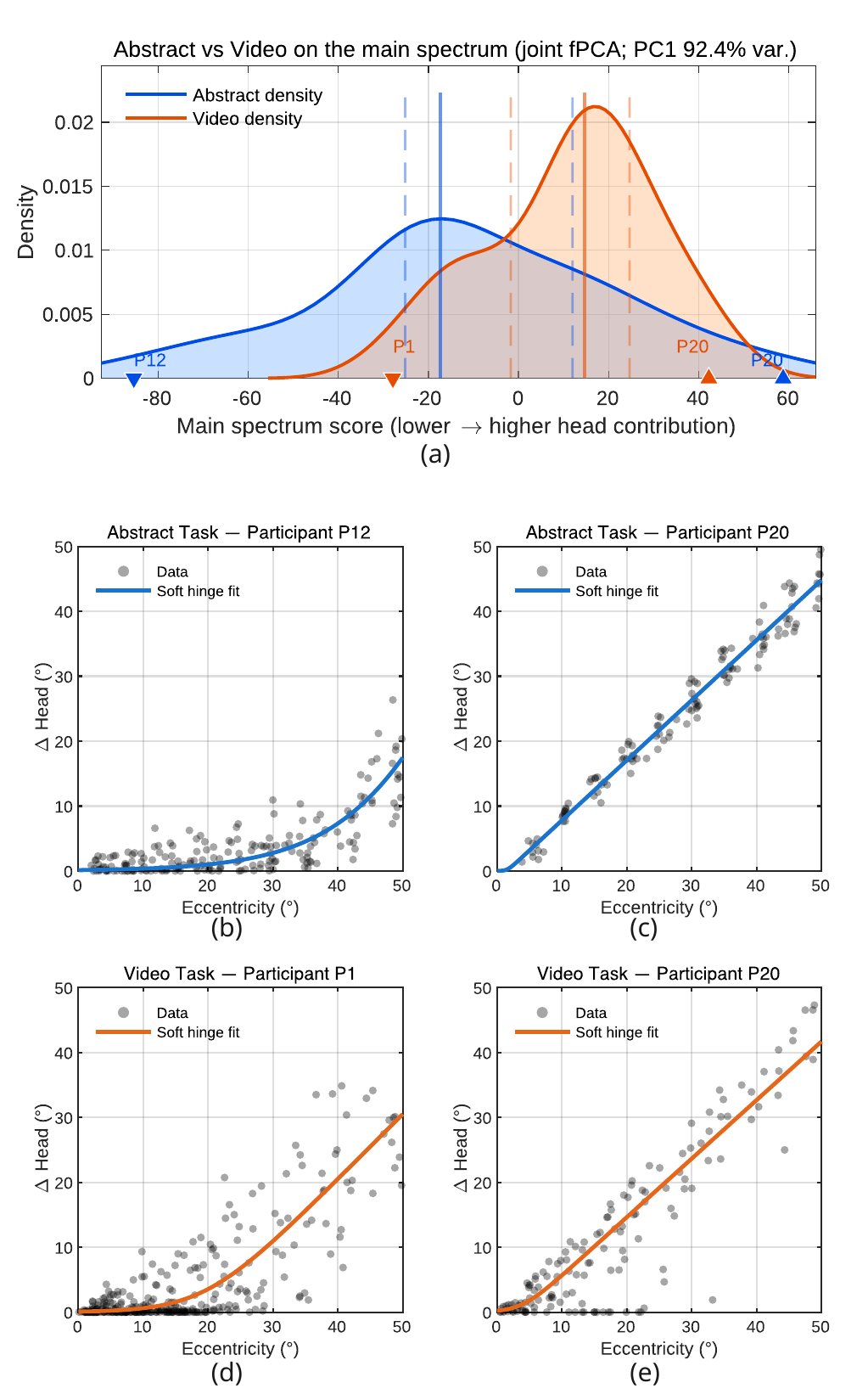}
    \caption{Main spectrum of variation across tasks (joint fPCA).
(a) Distribution of participant positions on the main spectrum (joint fPCA) for the Abstract task (blue) and the Video task (orange). Density curves show the overall distribution. Vertical lines mark the first quartile (dashed), median (solid), and third quartile (dashed) for each task. Triangles at the axis mark the extreme participants at the eye-mover and head-mover ends, annotated with their participant IDs. (b–e) Example soft hinge fits for extreme participants: P12 (Abstract, eye-mover), P20 (Abstract, head-mover), P1 (Video, eye-mover), P20 (Video, head-mover)}
\Description{Panel a shows density curves of Abstract (blue) and Video (orange) task distributions, with different quartiles. Panels b–e show model fits for extreme individuals, from strong eye-leaners to strong head-leaners.}
    \label{fig:joint_fpca_pc1}
\end{figure*}

To embed both tasks on a common axis, we conducted a joint functional PCA on the fitted soft-hinge curves from the Abstract and Video tasks, with each participant contributing two curves. The first component (PC1) explained 92.4\% of the variance and captured the dominant spectrum of coordination strategies from eye-mover to head-mover (\autoref{fig:joint_fpca_pc1}).

In the Abstract task ($N = 28$), participants spanned a wide range of spectrum scores, from –85.43 (P12, extreme eye-mover) to +58.79 (P20, extreme head-mover). The distribution was unimodal and approximately balanced (skewness = –0.094), with quartiles at –25.17 (Q1), –17.39 (median), and +11.99 (Q3). This widespread indicates that the Abstract task elicited both strong eye-mover and strong head-mover strategies.

In the Video task ($N = 28$), scores were more constrained, ranging from –27.90 (P1, eye-mover) to +42.28 (P20, head-mover). The distribution was unimodal but skewed toward the eye-mover side (skewness = –0.417), with quartiles at –1.73 (Q1), +14.62 (median), and +24.67 (Q3). Compared to the Abstract task, the Video task distribution was narrower and shifted toward intermediate-to-head-mover strategies.

Together, these results show that while both tasks spanned the full eye–head mover spectrum, they differed in emphasis: the Abstract task revealed a broader spread with more extreme eye-mover cases, whereas the Video task compressed participants toward the centre and head-mover side. Representative model fits for extreme individuals (P12, P20, and P1; \autoref{fig:joint_fpca_pc1}(b–e)) illustrate these differences. \newmarker{For illustration, we show participants at the extremes of the observed spectrum, as these cases most clearly convey the full range of individual differences.}

\subsubsection{Task Correlations}

\begin{figure}[tbh]
    \centering   \includegraphics[width=\linewidth]{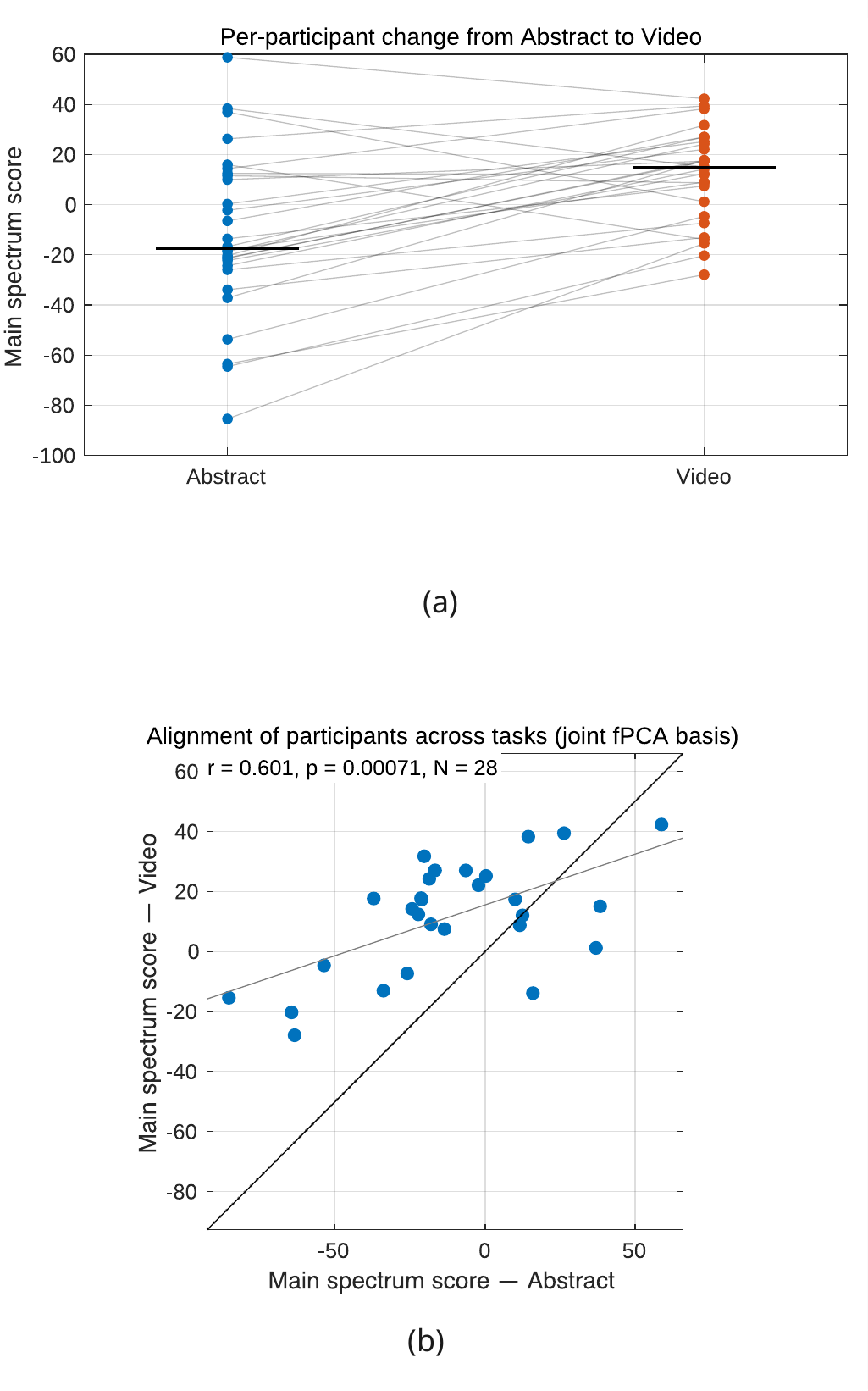}
    \caption{Participant alignment across tasks (joint fPCA).
(a) Scatterplot of participant scores on the main spectrum in Abstract vs Video tasks. A significant correlation ($r = 0.60$, $p < .001$, $N = 28$) indicates moderate alignment across tasks. (b) Slopegraph showing per-participant change, highlighting the diversity of shifts between Abstract and Video tasks.}
\Description{Panel a scatterplot shows a positive correlation between Abstract and Video spectrum scores. Panel b slopegraph shows each participant’s shift between tasks, with many moving upward toward more head involvement.}

    \label{fig:joint_fpca_corr}
\end{figure}

We next examined whether participants’ coordination strategies were consistent across the Abstract and Video tasks. Each participant’s PC1 scores from the joint fPCA space were compared directly between tasks.

As shown in \autoref{fig:joint_fpca_corr}(a), most participants shifted upward from Abstract to Video, indicating higher spectrum scores in the Video task. This shift reflects the narrower, head-mover–shifted distribution of Video scores reported earlier. However, the degree of change varied: while some participants moved closer to the centre, others retained their relative extremes.

Across participants, Abstract and Video PC1 scores were significantly correlated ($r = 0.601$, $p = 0.00071$, $N = 28$; \autoref{fig:joint_fpca_corr}b). This moderate-to-strong correlation indicates that individuals largely preserved their relative position on the eye–head mover spectrum across tasks.

\newnewmarker{With $N = 28$ participants, the study is sufficiently powered to detect meaningful, moderate-to-large cross-task consistency between Abstract and Video conditions. Specifically, a sensitivity-based power analysis (two-sided Pearson correlation, $\alpha = 0.05$) indicates 80\% power to detect correlations of $r \approx 0.51$ or larger. The observed effect ($r = 0.60$) lies well within this detectable range, corresponding to an achieved power of approximately 94\%.}

\subsection{Subjective Feedback}

\begin{figure*}[tbh]
    \centering   \includegraphics[width=1\linewidth]{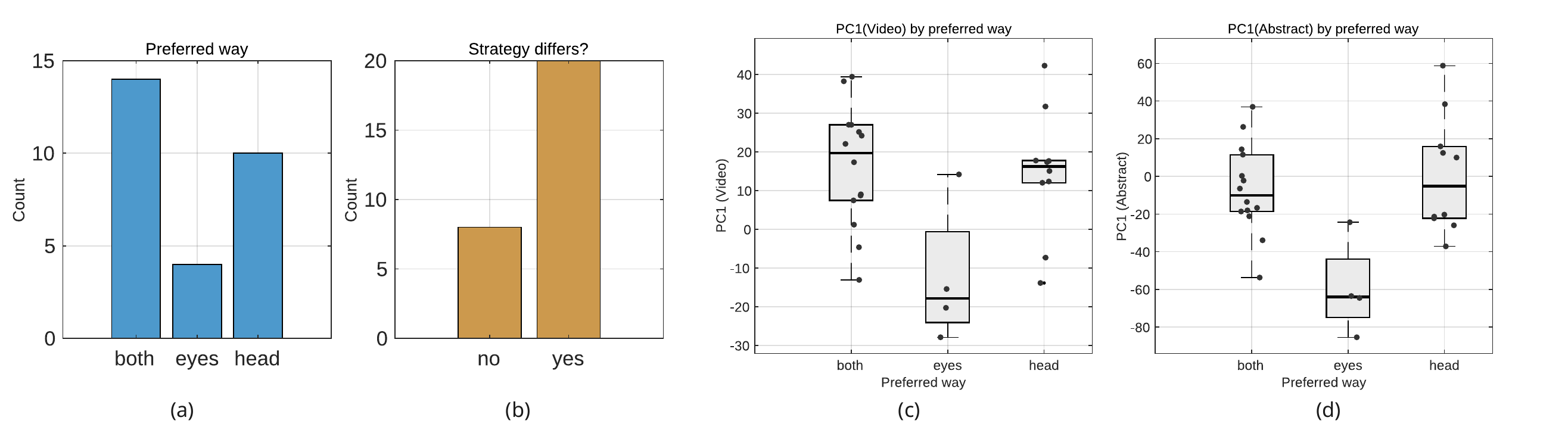}
    \caption{Subjective measures. (a-b) Awareness ratings for Abstract and Video tasks. (c-d) Bocplots of preferred ways of shifting gaze for Abstract and Video tasks.}
    \Description{Bar and boxplots showing self-reports. Awareness ratings are moderate and similar across tasks. Preferred strategies split between eyes, head, and both, with many participants reporting different strategies across tasks.}

    \label{fig:subjective}
\end{figure*}
% oee participant reported he had different strategies for left and right
Participants reported moderate awareness of their head movements in both tasks. On a 7-point Likert scale, awareness ratings were similar between the Abstract task (M = 4.13, SD = 1.15) and the Video task (M = 4.00, SD = 1.50). Normality of the paired differences was not satisfied ($p = .039$), and a Wilcoxon signed-rank test confirmed no significant difference between conditions (p = .756, median difference = 0).

When asked about their preferred way of shifting gaze, half of the participants (50.0\%) indicated that they used both eyes and head, while the remainder split into eye-leaning (14.3\%) and head-leaning (35.7\%) strategies, as shown in \autoref{fig:subjective}(a). Participants who reported using only their eyes also showed lower PC1 scores in both tasks (see \autoref{fig:subjective}(c-d)). By contrast, those who reported “head” or “both” did not cluster as clearly, with scores spanning a broader range across the spectrum.

Finally, most participants (71.4\%) reported that their strategy differed between the two tasks, while a smaller group (28.6\%) felt their approach remained consistent (see \autoref{fig:subjective}(b)).

\section{Discussion}
% revisit the 3 research questions
%  differences
%  prevelance
%  task 
% also propose the model method contribution
% model can be used later in...

Our study establishes the existence of a \textit{eye-head mover spectrum} in VR, capturing how individuals differ in the extent to which they involve the head when shifting gaze. Rather than splitting into discrete types, people spread along a continuous distribution that spans eye-leaning to head-leaning strategies. Most people exhibit different levels of mixed strategies between the extremes. This spectrum reflects user-specific preferences that were stable both within tasks and across tasks, while also shifting in expression depending on context. By introducing the \textit{eye-head mover spectrum}, we highlight a previously underexplored dimension of individual difference in VR that has direct implications for system design, adaptation, and inclusivity.

% Our work takes a first and fundamental step in establishing head movement tendencies as a new dimension of individual differences, comparable to handedness or eye dominance~\cite{prummer2025s}. This reframes design assumptions: rather than treating users as following a uniform strategy when shifting their gaze, VR systems should recognise and adapt to the diversity of coordination styles. 

\subsection{Individual Variation in Head Contribution (RQ1)}
% \paragraph{RQ1: How does head contribution to gaze shifts vary between individuals?}
Head involvement varied substantially across people. Some participants relied primarily on their eyes and delayed head movement until gaze shifts became large, while others engaged the head almost immediately, even for small eccentricities. These differences persisted even at amplitudes where head use is unavoidable: some participants still contributed relatively little, while others rotated their head much more. In the Video free-viewing task, this variation was clear at both small and large amplitudes, with contributions spanning ~5–15\textdegree{} at 20\textdegree{} and still diverging by more than 10\textdegree{} at 50\textdegree{}. The Abstract task magnified these contrasts: eye-leaning participants sometimes delayed head involvement until around 30\textdegree{} and rotated only modestly even at 50\textdegree{}, whereas head-leaning participants moved their head almost constantly, turning even at 5\textdegree{} and keeping their eyes near the centre of the field of view. Such contrasts underline that head contribution is not merely dictated by amplitude but represents a stable aspect of coordination strategy.

\subsection{Prevalence of Tendencies (RQ2)}

% \paragraph{RQ2: How prevalent are different tendencies in the population?}
Head-movement tendencies are widely prevalent across the population but not evenly distributed. The overall distribution was unimodal and continuous, with most participants clustering in an intermediate range that combined both eye and head movements. At the same time, the spectrum extended meaningfully in both directions: a non-trivial minority leaned strongly toward low head involvement, and another minority toward high head involvement. These ends are not isolated outliers but represent strategies that are prevalent enough to be considered in design. This continuous spectrum view helps explain why prior work \cite{fuller_head_1992} struggled to define categorical groups of “head movers” and “non-head movers”: the population is not split but instead spread smoothly across the range. 

% Recognising this has practical value, as it shifts the design challenge from addressing discrete types to supporting a continuum of strategies. 

\subsection{Consistency and Context Dependence (RQ3)}
% \paragraph{RQ3: Are individual tendencies in head contribution consistent across tasks?}
Individual strategies were consistent across tasks, but their expression shifted with context. Participants who moved their head more in the Abstract task also tended to do so in the Video task, meaning people largely kept their relative positions within the population distribution. At the same time, the two tasks produced different shapes of distribution. The Video task compressed behaviours toward higher head involvement, reflecting the demands of naturalistic free viewing of 360\textdegree{} videos. In contrast, the Abstract task polarised strategies toward both sides, revealing very strong eye-leaning and very strong head-leaning patterns. Within the Abstract task, these strategies were stable across repetitions, with no evidence of learning or drift across trial sets. 

These findings suggest that eye-head coordination reflects a user-specific preference that holds both within a task and across tasks, while still being modulated by contextual demands. Such modulation is likely broader than task type alone: factors such as dwell time, the nature of stimuli, or the presence of audio cues may all shape how individuals balance eye and head movements. Moreover, different forms of eye movement, such as smooth pursuit during continuous motion, could interact differently with head involvement than discrete gaze shifts. Together, these considerations point to a population distribution that not only captures the range of individual strategies but also their prevalence under varying contexts.

%     \item Other factors can affect the head/eye movement strategies: dwell time, stimuli type, audio cues…
%     \item test for different eye movements such as smooth pursuit, for different taskto derive a population distribution that reveals both the range and prevalence of different strategies.

% Our findings suggest that while large-scale naturalistic datasets can reveal population distributions, in practical adaptive systems, a short calibration task is essential to reliably estimate individual parameters. This calibration complements free-viewing, which by itself does not provide sufficient eccentricity coverage within short sessions
% we can also position the user study participants into the broad distribution of population from the dataset analysis using the same task.

% Our study establishes the existence of the eye-head mover spectrum in VR.

% \subsection{Task Context and Stability of Tendencies}

\subsection{Methodological Contributions}
The soft-hinge model serves not only as a descriptive tool for analysing eye-head coordination but also as a directly applicable representation. It avoids reliance on arbitrary thresholds and yields parameters that are both interpretable and robust to uneven or sparse data. These parameters can be incorporated into existing eye–head coordination models\cite{pan_head-eyek_2025,zhao_coordauth_2025}, bridging psychological insights with computational pipelines. The same formulation can, in principle, be used in adaptive VR/AR systems, offering a lightweight, user-specific profile for applications such as foveated rendering \cite{patney2016towards,  krajancich2023towards,pan_head-eyek_2025}, viewport alignment \cite{gottsacker_decoupled_2025,lee_patterns_2024}, or multi-user synchronisation \cite{bovo_cone_2022,maloney_talking_2020}. It also allows for implicit updating during use, enabling profiles to adapt dynamically over time. While we do not demonstrate such applications here, the model provides both a solid analytic basis and a pathway toward practical integration.

% cues that make joint attention less transparent \cite{bovo_cone_2022,maloney_talking_2020}. This variation also affects the alignment of what collaborators actually see, as differences in head movement change how closely their viewpoints match when looking at the same content \cite{gottsacker_decoupled_2025,lee_patterns_2024}. Finally, at the system level, modelling variation in head contribution improves gaze prediction \cite{hu_sgaze_2019} and adaptive viewport streaming \cite{corbillon2017viewport, hosseini2016adaptive}, and can support more efficient foveated rendering \cite{patney2016towards,  krajancich2023towards,pan_head-eyek_2025}.

We further extend from individual profiles to a population perspective. Individual characteristics are often difficult to establish convincingly in isolation, as they risk being dismissed as noise or idiosyncrasy. By embedding profiles within a distribution, we reveal both the range and prevalence of strategies, making it clear which tendencies are common and which are rare. This population view enables individual differences to be interpreted as positions on a shared continuum rather than as outliers, providing a reproducible basis for systematic comparison across tasks and studies.

%     \item the model can be updated during use implicitly. 
%     \item How can these individual differences be used in eye-head coordination models? the model can be directly incorporated into the eye head coordination models
% can also e directly used for adaptive systems

%  we also provide a perspective to look at individual differences 
\subsection{Application}

% demonstrate the effectiveness of the model to show different aspects for each application:
% \begin{itemize}
%     \item immersive storytelling: align different individuals 
%     \item kuper belt: different for individuals
%     \item layout: use less neck movements
% \end{itemize}
% application area:
% foveated rendering 
% multi user experience alignment
% virtual characters
How can our findings of different individual behaviours help? This understanding can help VR systems in various applications relevant to perceiving, interacting, and system optimisations. We describe some of the use cases here. 

In foveated rendering, eye-head tendencies determine both the size and update rate of high-resolution regions. Eye-leaning users sustain larger eccentricities, so they benefit from a broader foveal area that updates less frequently. By contrast, head-leaning users keep their gaze nearer the centre, allowing for a smaller high-resolution area but requiring more frequent updates as the head shifts the viewport. Similar advantages extend to viewport prediction and adaptive streaming, where incorporating such profiles improves both efficiency and accuracy.

Interface layouts can be adapted for comfort. For head-leaning users, menus and annotations can be positioned more widely since head rotation is naturally engaged, whereas for eye-leaning users, content should be placed closer to the centre to minimise neck effort.

For some novel interaction techniques, such as Kuiper Belt\cite{choi2022kuiper}, the effective “belt” range can be tuned individually.

In collaborative VR, differences in head involvement affect the clarity of orientation cues. Head-leaning users provide overt signals that others can easily follow, while eye-leaning users are subtler, requiring systems to support cues more actively. 

In immersive storytelling, profiles could help align perspectives or viewports across diverse audiences. For avatars and characters, distinct coordination styles enrich realism and social presence.

Finally, the modelling framework supports personalisation over time. Because profiles can be updated implicitly during use, systems need not assume fixed strategies. Instead, they can adapt dynamically as individual behaviours evolve with context, comfort, or task demands.
% This opens opportunities for personalization in foveated rendering,
% \subsection{Future Work}
% **blend all the future work into discussions
% \begin{itemize}
%     \itemthe model can be updated during use implicitly. 
%     \item How can these individual differences be used in eye-head coordination models?
%     \item Our analysis and model focus on the horizontal direction.
%     \item Other factors can affect the head/eye movement strategies: dwell time, stimuli type, audio cues…
%     \item test for different eye movements such as smooth pursuit, for different taskto derive a population distribution that reveals both the range and prevalence of different strategies.
    
% \end{itemize}

% \subsection{Boarder Perspective}
% establishes head-eye tebdebct as a new dimension of individual differences, comparable to handedness or vision.
% reframes desgin: systems should not assume uniform strategies, but account for diversity.
% reframes gaze interaction in VR as not just eye-driven but a eye-head continuum that matters for design.
% \textcolor{red}{Not sure if this is right...}
% Our work takes a first and fundamental step in establishing head movement tendencies as a new dimension of individual differences, comparable to handedness or eye dominance~\cite{prummer2025s}. This reframes design assumptions: rather than treating users as following a uniform strategy when shifting their gaze, VR systems should recognise and adapt to the diversity of coordination styles. 

\subsection{Limitations}

Several limitations should be acknowledged. \deletemarker{ Our analysis focused on horizontal gaze and head movements, while vertical contributions and full 3D coordination remain unexplored.} The user study and dataset revealed consistent patterns, but the tasks we examined (abstract target selection and 360\textdegree{} video viewing) cover only a subset of VR contexts. Other factors, such as stimulus type, dwell time, or the presence of audio cues, may also shape strategies. Our user study also had limited gender balance, which may have influenced the observed distribution of behaviours. Additionally, the modelling framework offers one useful representation, yet we do not claim it to be the definitive approach. It provides interpretable profiles but does not yet capture dynamic changes in real time.

\newmarker{Our findings should also be interpreted within the empirical scope of the data and tasks we analysed. First, our population-level analysis draws on a single large-scale dataset (D-SAV360). As far as we know, this is currently the largest publicly available dataset that includes both head and eye tracking for natural free viewing in VR, and therefore provides the strongest existing basis for modelling population tendencies at scale. Nevertheless, transferability to additional large datasets remains an open direction for future work. Second, our analysis is limited to horizontal coordination within a ±50° range. In the naturalistic video dataset, vertical gaze shifts are both sparse and restricted to a narrow amplitude range, and our user study was likewise designed horizontally, leaving insufficient vertical data for reliable modelling. Extending the spectrum to full 3D coordination is therefore an important next step. Third, amplitudes beyond 50° increasingly recruit the torso, shifting the coordination problem from eye–head to eye–head–torso dynamics. Torso involvement substantially changes the mechanical load and redistributes rotation across joints, which would meaningfully alter head contributions; the current model is not intended for this expanded regime.}

\newmarker{In addition, the viewing conditions only represent standing posture, monoscopic 360° videos, and static free-viewing, and reflect only a subset of VR experiences. Other contexts such as locomotion, interaction, or stereoscopic videos, may engage head movement differently. Likewise, our datasets do not include broader demographic, perceptual, or personality measures, which leaves open how such factors might relate to an individual’s position on the Eye--Head mover spectrum. Finally, both datasets were collected using VR headsets, which introduce additional mass and inertia. While existing literature does not establish a specific directional bias for the devices we used (Vive Pro Eye and Quest Pro), such ergonomic factors can influence the absolute level or timing of head involvement. Because all participants experienced the same hardware constraints within each dataset, the spectrum we report reflects relative individual differences under shared device ergonomics, rather than a device-independent biomechanical baseline.}

% what does this tell us

% who can benefit from this

% how can this help 

\subsection{\newnewmarker{Implications for Accessibility and Inclusive Design}}
\newnewmarker{The eye–head mover spectrum also has implications for accessibility. Many VR/AR interactions implicitly assume a typical eye–head coordination pattern, yet our results show that coordination strategies vary continuously across individuals. In both the large-scale dataset (Figure \ref{fig:fpca_360_distribution}) and the user study (Figure \ref{fig:fpca_us_distribution}), participants spanned wide and continuous ranges along the spectrum, with a non-trivial fraction occupying positions far from the central tendency. Designs that favour specific eye–head movement behaviours may therefore reduce comfort or efficiency for users whose natural coordination lies toward the extremes.}

\newnewmarker{More broadly, eye–head coordination tendency can be viewed as a fundamental user characteristic that, similar to handedness or reading speed, can systematically influence behaviour across VR interactions. Treating this behavioural heterogeneity as a stable dimension of user behaviour, rather than as noise or inefficiency, supports designs that accommodate multiple coordination strategies and helps ensure that immersive systems remain usable and comfortable across a broader range of users.}

\section{Conclusion}

% Key takeaways
We introduced eye-head movement tendencies as a new dimension of individual difference in VR. We proposed a user-specific model of how head contribution scales with gaze shift amplitude and revealed a continuous spectrum from eye-leaning to head-leaning strategies. Analysis of a large dataset and a controlled user study showed that these strategies are stable within individuals yet shift with task context. These findings establish both the existence and prevalence of such tendencies and \newnewmarker{represents a fundamental behavioural heterogeneity that can influence performance and experience across many VR interactions.}
% lol hello
\begin{acks} 

This work was supported by the European Research Council (ERC) under the European Union’s Horizon 2020 research and innovation program (Grant No. 101021229 GEMINI: Gaze and Eye Movement in Interaction).

\end{acks}

\bibliographystyle{ACM-Reference-Format}
\bibliography{sample-base}
\end{document}